\documentclass[12pt]{article}
\usepackage{graphicx}
\usepackage{lscape}
\usepackage{citesort}

\begin{document}

\def\ds{\displaystyle}
\def\beq{\begin{equation}}
\def\eeq{\end{equation}}
\def\bea{\begin{eqnarray}}
\def\eea{\end{eqnarray}}
\def\beeq{\begin{eqnarray}}
\def\eeeq{\end{eqnarray}}
\def\ve{\vert}
\def\vel{\left|}
\def\ver{\right|}
\def\nnb{\nonumber}
\def\ga{\left(}
\def\dr{\right)}
\def\aga{\left\{}
\def\adr{\right\}}
\def\lla{\left<}
\def\rra{\right>}
\def\rar{\rightarrow}
\def\nnb{\nonumber}
\def\la{\langle}
\def\ra{\rangle}
\def\ba{\begin{array}}
\def\ea{\end{array}}
\def\tr{\mbox{Tr}}
\def\ssp{{\Sigma^{*+}}}
\def\sso{{\Sigma^{*0}}}
\def\ssm{{\Sigma^{*-}}}
\def\xis0{{\Xi^{*0}}}
\def\xism{{\Xi^{*-}}}
\def\qs{\la \bar s s \ra}
\def\qu{\la \bar u u \ra}
\def\qd{\la \bar d d \ra}
\def\qq{\la \bar q q \ra}
\def\gGgG{\la g^2 G^2 \ra}
\def\q{\gamma_5 \not\!q}
\def\x{\gamma_5 \not\!x}
\def\g5{\gamma_5}
\def\sb{S_Q^{cf}}
\def\sd{S_d^{be}}
\def\su{S_u^{ad}}
\def\ss{S_s^{??}}
\def\sbp{{S}_Q^{'cf}}
\def\sdp{{S}_d^{'be}}
\def\sup{{S}_u^{'ad}}
\def\ssp{{S}_s^{'??}}
\def\sig{\sigma_{\mu \nu} \gamma_5 p^\mu q^\nu}
\def\fo{f_0(\frac{s_0}{M^2})}
\def\ffi{f_1(\frac{s_0}{M^2})}
\def\fii{f_2(\frac{s_0}{M^2})}
\def\O{{\cal O}}
\def\sl{{\Sigma^0 \Lambda}}
\def\es{\!\!\! &=& \!\!\!}
\def\ap{\!\!\! &\approx& \!\!\!}
\def\ar{&+& \!\!\!}
\def\ek{&-& \!\!\!}
\def\kek{\!\!\!&-& \!\!\!}
\def\cp{&\times& \!\!\!}
\def\se{\!\!\! &\simeq& \!\!\!}
\def\eqv{&\equiv& \!\!\!}
\def\kpm{&\pm& \!\!\!}
\def\kmp{&\mp& \!\!\!}

% .........................................................

\def\simlt{\stackrel{<}{{}_\sim}}
\def\simgt{\stackrel{>}{{}_\sim}}

% .........................................................

\title{
         {\Large
                 {\bf
Double--lepton polarization asymmetries in the 
Exclusive $B \rar \rho \ell^+ \ell^-$ decay 
beyond the Standard Model 
                 }
         }
      }

\author{\vspace{1cm}\\
{\small T. M. Aliev\,$^{(1)}$ \thanks
{e-mail: taliev@metu.edu.tr}
\,\,,
V. Bashiry% \thanks
%{e-mail: e114288@metu.edu.tr}
\,\,,
M. Savc{\i} \thanks
{e-mail: savci@metu.edu.tr}} \\
{\small Physics Department, Middle East Technical University,
06531 Ankara, Turkey} \\
{\small (1) Permanent Adress: Institute of Physics, Baku, Azerbaycan} }

\date{}

\begin{titlepage}
\maketitle
\thispagestyle{empty}

\begin{abstract}
The double--lepton polarization asymmetries in $B \rar
\rho \ell^+ \ell^-$ decay is analyzed in a model independent framework.
The general expressions for nine double--polarization asymmetries are 
calculated. It is shown that the study of the double--lepton polarization 
asymmetries proves to be very useful tool in looking for new physics 
beyond the standard model.
\end{abstract}

%\vspace{1cm}
~~~PACS numbers: 13.20.He, 12.60.--i, 13.88.+e
\end{titlepage}

\section{Introduction}
Rare $B$ meson decays, induced by flavor changing neutral current (FCNC)
$b \rar s(d) \ell^+ \ell^-$ transitions constitute one of the most important
class of decays for testing the gauge structure of the Standard Model (SM). 
These decays which are forbidden in the SM at tree level, occur at loop level 
and provide insight to check the predictions of the SM at quantum level. 
Moreover, these decays are also 
quite sensitive to the existence of new physics beyond the SM, since
new particles running at loops can give contribution to these decays. The
new physics manifests itself in rare decays in two different ways; one via
modification of the existing Wilson coefficients in the SM, or through the
introduction of some new operators with new coefficients which are absent in
the SM. Some of the most important exclusive FCNC decays governed by $b \rar
s(d)$ transition at quark level are $B \rar K^\ast \gamma$
and $B \rar (\pi,\rho,K,K^\ast) \ell^+ \ell^-$ decays. The decays of the
kind $B \rar M \ell^+ \ell^-$, where $M$ stands for pseudoscalar or vector
mesons, enable the investigation of the experimental observables, such as,
lepton pair forward--backward (FB) asymmetry, lepton polarizations, etc. 
One of the most efficient ways in looking for new physics beyond the SM is
the measurement of lepton polarization in the decays. Polarization of a
single lepton has been studied in  $B \rar K^\ast \ell^+ \ell^-$
\cite{R7001}, $B \rar X_s \ell^+ \ell^-$ \cite{R7002,R7003}, $B \rar K
\ell^+ \ell^-$ \cite{R7004},  $B \rar \pi (\rho) \ell^+
\ell^-$\cite{R7005,R7006} and $B \rar \ell^+ \ell^- \gamma$ \cite{R7007} 
decays in detail in fitting the parameters of the SM and set constraints 
on new physics beyond the SM. Moreover, as has already
been pointed out in \cite{R7008}, some of the single lepton polarization
asymmetries might be quite small to be observed and might not provide
sufficient number of observables in checking the structure of the effective
Hamiltonian. By taking both lepton polarizations into account
simultaneously, maximum number of independent observables
are constructed. It is clear that, 
measurement of many more observables which would be useful in further
improvement of the parameters of the SM probing new physics beyond the SM.
It should be noted here that both lepton polarizations in the $B \rar K^\ast
\tau^+ \tau^-$ and $B \rar K \ell^+ \ell^-$ decays are studied in
\cite{R7009} and  \cite{R7010}, respectively.
The decays of $B$ mesons
induced by the $b \rar d \ell^+ \ell^-$ transition are promising in looking
for CP violation since the CKM factors $V_{tb} V_{td}^\ast$, 
$V_{ub} V_{ud}^\ast$ and $V_{cb} V_{cd}^\ast$ in the SM are all of the same
order. For this reason CP violation is much more considerable in the decays
induced by $b \rar d$ transition. So, study of the exclusive decays $B_d \rar
(\pi,\rho,\eta) \ell^+ \ell^-$ are quite promising for the confirmation of
the CP violation and these decays have extensively been investigated in the
SM \cite{R7011} and beyond \cite{R7012}.   

The aim of the present work is to study the double--lepton polarization
asymmetries in the exclusive $B \rar \rho \ell^+ \ell^-$ decay in a model
independent way, including all possible forms of interactions into the
effective Hamiltonian. Moreover, we study the correlation between the 
double--lepton polarization asymmetries and the branching ratio of the $B \rar
\rho \ell^+ \ell^-$ decay, in order to find such regions of new Wilson
coefficients in which the branching ratio (as well as single--lepton
polarization) coincides with the SM prediction while the double--lepton
polarization asymmetries do not. It is clear that if such a region of the
new Wilson coefficients exists it is an indication of the fact that new
physics beyond the SM can be established by measurement of the
double--lepton polarizations only. Note that the double--lepton
polarizations in the $B \rar K \ell^+ \ell^-$ and $B \rar \ell^+ \ell^-
\gamma$ decays are studied in \cite{R7013} and \cite{R7014} in detail. 

The paper is organized as follows. In section 2, using a general form
of the effective Hamiltonian, we obtain the matrix element
in terms of the form factors of the $B \rar \rho$ transition. In section 3
we derive the analytical results for the nine double--lepton polarization
asymmetries. Last section is devoted to the numerical analysis, discussion and
conclusions.

\section{Double lepton polarization asymmetries in 
$B \rar \rho \ell^+ \ell^-$ decay}

In this section we calculate the double lepton polarizations 
using a general form of the effective Hamiltonian. 
The $B \rar \rho \ell^+ \ell^-$ process is governed
by $b \rar d \ell^+ \ell^-$ transition at quark level. The matrix element
for the $b \rar d \ell^+ \ell^-$ transition can be written in terms of the
twelve model independent four--Fermi interactions in the following form:
\bea
\label{e7001}
{\cal H}_{eff} \es \frac{G_F\alpha}{\sqrt{2} \pi}
 V_{td}V_{tb}^\ast
\Bigg\{ C_{SL} \, \bar{d}_R i \sigma_{\mu\nu} \frac{q^\nu}{q^2}\, b_L
\, \bar \ell \gamma^\mu \ell + C_{BR}\, \bar{d}_L i \sigma_{\mu\nu}
\frac{q^\nu}{q^2} \, b_R \, \bar \ell \gamma^\mu \ell \nnb \\
\ar C_{LL}^{tot}\, \bar d_L \gamma_\mu b_L \,\bar \ell_L \gamma^\mu \ell_L +
C_{LR}^{tot} \,\bar d_L \gamma_\mu b_L \, \bar \ell_R \gamma^\mu \ell_R +
C_{RL} \,\bar d_R \gamma_\mu b_R \,\bar \ell_L \gamma^\mu \ell_L \nnb \\
\ar C_{RR} \,\bar d_R \gamma_\mu b_R \, \bar \ell_R \gamma^\mu \ell_R +
C_{LRLR} \, \bar d_L b_R \,\bar \ell_L \ell_R +
C_{RLLR} \,\bar d_R b_L \,\bar \ell_L \ell_R \nnb \\
\ar C_{LRRL} \,\bar d_L b_R \,\bar \ell_R \ell_L +
C_{RLRL} \,\bar d_R b_L \,\bar \ell_R \ell_L+
C_T\, \bar d \sigma_{\mu\nu} b \,\bar \ell \sigma^{\mu\nu}\ell \nnb \\
\ar i C_{TE}\,\epsilon^{\mu\nu\alpha\beta} \bar d \sigma_{\mu\nu} b \,
\bar \ell \sigma_{\alpha\beta} \ell  \Bigg\}~,
\eea
where
\bea
d_L = \frac{1-\gamma_5}{2}d ~,~~~~~~ d_R = \frac{1+\gamma_5}{2}d\nnb~,
\eea
$C_X$ are the coefficients of the four--Fermi interactions and $q$ is the
momentum transfer. Among all these Wilson coefficients, several already
exits in the SM. Indeed,
the first two coefficients in Eq. (\ref{e7001}), $C_{SL}$ and $C_{BR}$, are 
the nonlocal Fermi interactions, which correspond to $-2 m_s C_7^{eff}$ and 
$-2 m_b C_7^{eff}$ in the SM, respectively. The next four terms
with coefficients $C_{LL}$, $C_{LR}$, $C_{RL}$ and $C_{RR}$ are the
vector type interactions. Two of these vector
interactions containing $C_{LL}^{tot}$ and $C_{LR}^{tot}$ do already
exist in the SM in the form $(C_9^{eff}-C_{10})$ and $(C_9^{eff}+C_{10})$.
Therefore, $C_{LL}^{tot}$ and $C_{LR}^{tot}$ can be written as
\bea
C_{LL}^{tot} &=& C_9^{eff} - C_{10} + C_{LL}~, \nnb \\
C_{LR}^{tot} &=& C_9^{eff} + C_{10} + C_{LR}~, \nnb
\eea
where $C_{LL}$ and $C_{LR}$ describe the contributions of the new physics.
The terms with
coefficients $C_{LRLR}$, $C_{RLLR}$, $C_{LRRL}$ and $C_{RLRL}$ describe
the scalar type interactions. The remaining last two terms lead by the
coefficients $C_T$ and $C_{TE}$, obviously, describe the tensor type
interactions.

It should be noted here that, in further analysis we will assume that all new
Wilson coefficients are real, as is the case in the SM, while only
$C_9^{eff}$ contains imaginary part and it is parametrized in the following
form
\bea
\label{e7002}
C_9^{eff} = \xi_1 + \lambda_u \xi_2~,
\eea
where
\bea
\lambda_u = \frac{V_{ub} V_{ud}^\ast}{V_{tb} V_{td}^\ast} \nnb~,
\eea
and
\bea
\label{e7003} 
\xi_1 \es 4.128  + 0.138 \omega(\hat{s}) + g(\hat{m}_c,\hat{s})
C_0(\hat{m}_b) 
- \frac{1}{2} g(\hat{m}_d,\hat{s}) (C_3 + C_4) \nnb \\
\ek \frac{1}{2}
g(\hat{m}_b,\hat{s}) (4 C_3 + 4 C_4 + 3C_5 + C_6)
+ \frac{2}{9} (3 C_3 + C_4 + 3C_5 + C_6)~,\nnb \\
\xi_2 \es [g(\hat{m}_c,\hat{s}) - g(\hat{m}_u,\hat{s})](3 C_1 + C_2)~,
\eea 
where $\hat{m}_q = m_q/m_b$, $\hat{s}=q^2$, $C_0(\mu)=3 C_1 + C_2 + 3 C_3 + 
C_4 + 3 C_5 + C_6$, and
\bea
\label{e7004}
\omega(\hat{s}) \es -\frac{2}{9} \pi^2 -\frac{4}{3} Li_2(\hat{s})-
\frac{2}{3} \ln (\hat{s}) \ln(1-\hat{s}) -
\frac{5+4\hat{s}}{3(1+2\hat{s})} \ln(1-\hat{s}) \nnb \\
\ek \frac{2 \hat{s}(1+\hat{s})(1-2\hat{s})}{3(1-\hat{s})^2(1+2\hat{s})}
\ln (\hat{s}) + \frac{5+9 \hat{s}-6 \hat{s}^2}{3(1-\hat{s})(1+2\hat{s})}~,
\eea
represents the $O(\alpha_s)$ correction coming from one gluon exchange
in the matrix element of the operator ${\cal O}_9$ \cite{R7015}, while the 
function $g(\hat{m}_q,\hat{s})$ represents one--loop corrections to the 
four--quark operators $O_1$--$O_6$ \cite{R7016}, whose form is
\bea
\label{e7005}
\lefteqn{
g(\hat{m}_q,\hat{s}) = -\frac{8}{9} \ln (\hat{m}_q) + \frac{8}{27} + 
\frac{4}{9} y_q - \frac{2}{9} (2+y_q)} \nnb \\
\ek \sqrt{\vel 1-y_q \ver} \Bigg\{ \theta(1-y_q)\Bigg[\ln \Bigg(\frac{1+\sqrt{1-y_q}}
{1-\sqrt{1-y_q}}\Bigg) -i\pi \Bigg] + \theta(y_q-1) 
\arctan \Bigg( \frac{1}{\sqrt{y_q-1}}\Bigg) \Bigg\}~,
\eea
where $y_q=4 \hat{m}_q^2/\hat{s}$.

In addition to the short distance contributions, $B\rar X_d \ell^+ \ell^-$
decay also receives long distance contributions, which have their origin in
the real $\bar{u}u$, $\bar{d}d$ and $\bar{c}c$ intermediate states, i.e.,
$\rho$, $\omega$ and $J/\psi$ family. There are four different approaches in
taking long distance contributions into consideration: a) HQET based
approach \cite{R7017}, b) AMM approach \cite{R7018}, c) LSW
approach \cite{R7019}, and d) KS approach \cite{R7020}. In the present work we
choose the AMM approach, in which these resonance contributions are
parametrized using the Breit--Wigner form for the resonant states. The
effective coefficient $C_9^{eff}$ including the $\rho$, $\omega$ and
$J/\psi$ resonances are defined as
\bea
\label{e7006}
C_9^{eff} = C_9(\mu) + Y_{res}(\hat{s})~,
\eea
where
\bea
\label{e7007}
Y_{res} \es -\frac{3\pi}{\alpha^2} \Bigg\{ \ga C^{(0)} (\mu) + \lambda_u
\left[3 C_1(\mu)+ C_2(\mu) \right] \dr 
\sum_{V_i=\psi} K_i \frac{\Gamma(V_i \rar \ell^+\ell^-) M_{V_i}}
{M_{V_i}^2-q^2-iM_{V_i}\Gamma_{V_i}} \nnb \\
\ek \lambda_u g(\hat{m}_u,\hat{s}) \left[3 C_1(\mu)+ C_2(\mu) \right] 
\sum_{V_i=\rho,\omega} \frac{\Gamma(V_i \rar \ell^+\ell^-) M_{V_i}} 
{M_{V_i}^2-q^2-iM_{V_i}\Gamma_{V_i}} \Bigg\}~.
\eea
The phenomenological factor $K_i$ has the universal value for the inclusive 
$B \rar X_{s(d)}\ell^+ \ell^-$ decay $K_i\simeq 2.3$ \cite{R7021}, which we use 
in our calculations.

The decay amplitude for the exclusive $B \rar \rho \ell^+ \ell^-$ decay is 
obtained from the matrix element of the effective Hamiltonian in 
Eq. (\ref{e7001}) over $B$ and $\rho$ meson states, which can be parametrized 
in terms of various form factors. It follows from (\ref{e7001}) that, the 
following matrix elements  
\bea
&&\lla \rho\vel \bar d \gamma_\mu (1 \pm \gamma_5) 
b \ver B \rra~,\nnb \\
&&\lla \rho \vel \bar d i\sigma_{\mu\nu} q^\nu  
(1 \pm \gamma_5) b \ver B \rra~, \nnb \\
&&\lla \rho \vel \bar d (1 \pm \gamma_5) b 
\ver B \rra~, \nnb \\
&&\lla \rho \vel \bar d \sigma_{\mu\nu} b
\ver B \rra~, \nnb
\eea
are needed in obtaining the decay amplitude of the $B \rar \rho \ell^+ \ell^-$ decay. 
These matrix elements are defined as follows:
\bea
\lefteqn{
\label{e7008}
\lla \rho(p_{\rho},\varepsilon) \vel \bar d \gamma_\mu 
(1 \pm \gamma_5) b \ver B(p_B) \rra =} \nnb \\
&&- \epsilon_{\mu\nu\lambda\sigma} \varepsilon^{\ast\nu} p_{\rho}^\lambda q^\sigma
\frac{2 V(q^2)}{m_B+m_{\rho}} \pm i \varepsilon_\mu^\ast (m_B+m_{\rho})   
A_1(q^2) \\
&&\mp i (p_B + p_{\rho})_\mu (\varepsilon^\ast q)
\frac{A_2(q^2)}{m_B+m_{\rho}}
\mp i q_\mu \frac{2 m_{\rho}}{q^2} (\varepsilon^\ast q)
\left[A_3(q^2)-A_0(q^2)\right]~,  \nnb \\  \nnb \\
\lefteqn{
\label{e7009}
\lla \rho(p_{\rho},\varepsilon) \vel \bar d i \sigma_{\mu\nu} q^\nu
(1 \pm \gamma_5) b \ver B(p_B) \rra =} \nnb \\
&&4 \epsilon_{\mu\nu\lambda\sigma} 
\varepsilon^{\ast\nu} p_{\rho}^\lambda q^\sigma
T_1(q^2) \pm 2 i \left[ \varepsilon_\mu^\ast (m_B^2-m_{\rho}^2) -
(p_B + p_{\rho})_\mu (\varepsilon^\ast q) \right] T_2(q^2) \\
&&\pm 2 i (\varepsilon^\ast q) \left[ q_\mu -
(p_B + p_{\rho})_\mu \frac{q^2}{m_B^2-m_{\rho}^2} \right] 
T_3(q^2)~, \nnb \\  \nnb \\ 
\lefteqn{
\label{e7010}
\lla \rho(p_{\rho},\varepsilon) \vel \bar d \sigma_{\mu\nu} 
 b \ver B(p_B) \rra =} \nnb \\
&&i \epsilon_{\mu\nu\lambda\sigma}  \Bigg\{ - 2 T_1(q^2)
{\varepsilon^\ast}^\lambda (p_B + p_{\rho})^\sigma +
\frac{2}{q^2} (m_B^2-m_{\rho}^2) \Big[ T_1(q^2) - T_2(q^2) \Big] 
{\varepsilon^\ast}^\lambda q^\sigma \\
&&- \frac{4}{q^2} \Bigg[ T_1(q^2) - T_2(q^2) - \frac{q^2}{m_B^2-m_{\rho}^2} 
T_3(q^2) \Bigg] (\varepsilon^\ast q) p_{\rho}^\lambda q^\sigma \Bigg\}~. \nnb 
\eea
where $q = p_B-p_{\rho}$ is the momentum transfer and $\varepsilon$ is the
polarization vector of $\rho$ meson. 

Note that the matrix element 
\bea
\lla \rho(p_{\rho},\varepsilon) \vel \bar d
\sigma_{\mu\nu} \gamma_5 b \ver B(p_B) \rra \nnb
\eea
can easily be obtained from (\ref{e7010})by using the identity
\bea
\sigma_{\alpha\beta} = - \frac{i}{2} \epsilon_{\alpha\beta\rho\sigma}
\sigma^{\rho\sigma} \gamma_5~. \nnb
\eea 
In order to ensure finiteness of (\ref{e7008}) and (\ref{e7010}) at $q^2=0$, 
we assume that $A_3(q^2=0) = A_0(q^2=0)$ and $T_1(q^2=0) = T_2(q^2=0)$.
The matrix element $\lla \rho \vel \bar d (1 \pm \gamma_5 ) b \ver B \rra$
can be calculated by contracting both sides of Eq. (\ref{e7008}) 
with $q^\mu$ and using equation of motion. Neglecting the mass of the 
$d$ quark we get
\bea
\label{e7011}
\lla \rho(p_{\rho},\varepsilon) \vel \bar d (1 \pm \gamma_5) b \ver
B(p_B) \rra =
\frac{1}{m_b} \Big[ \mp 2i m_{\rho} (\varepsilon^\ast q)
A_0(q^2)\Big]~.
\eea
In deriving Eq. (\ref{e7011}) we have used the relationship
\bea
2 m_{\rho} A_3(q^2) = (m_B+m_{\rho}) A_1(q^2) -
(m_B-m_{\rho}) A_2(q^2)~, \nnb 
\eea
which follows from the equations of motion.

Using the definition of the form factors, as given above, the amplitude of
the  $B \rar \rho \ell^+ \ell^-$ decay can be written as 
\bea
\lefteqn{
\label{e7012}
{\cal M}(B\rightarrow \rho \ell^{+}\ell^{-}) =
\frac{G \alpha}{4 \sqrt{2} \pi} V_{tb} V_{td}^\ast }\nnb \\
&&\times \Bigg\{
\bar \ell \gamma^\mu(1-\gamma_5) \ell \, \Big[
-2 A_1 \epsilon_{\mu\nu\lambda\sigma} \varepsilon^{\ast\nu}
p_{\rho}^\lambda q^\sigma
 -i B_1 \varepsilon_\mu^\ast
+ i B_2 (\varepsilon^\ast q) (p_B+p_{\rho})_\mu
+ i B_3 (\varepsilon^\ast q) q_\mu  \Big] \nnb \\
&&+ \bar \ell \gamma^\mu(1+\gamma_5) \ell \, \Big[
-2 C_1 \epsilon_{\mu\nu\lambda\sigma} \varepsilon^{\ast\nu}
p_{\rho}^\lambda q^\sigma
 -i D_1 \varepsilon_\mu^\ast    
+ i D_2 (\varepsilon^\ast q) (p_B+p_{\rho})_\mu
+ i D_3 (\varepsilon^\ast q) q_\mu  \Big] \nnb \\
&&+\bar \ell (1-\gamma_5) \ell \Big[ i B_4 (\varepsilon^\ast
q)\Big]
+ \bar \ell (1+\gamma_5) \ell \Big[ i B_5 (\varepsilon^\ast
q)\Big]  \nnb \\
&&+4 \bar \ell \sigma^{\mu\nu}  \ell \Big( i C_T \epsilon_{\mu\nu\lambda\sigma}
\Big) \Big[ -2 T_1 {\varepsilon^\ast}^\lambda (p_B+p_{\rho})^\sigma +
B_6 {\varepsilon^\ast}^\lambda q^\sigma -
B_7 (\varepsilon^\ast q) {p_{\rho}}^\lambda q^\sigma \Big] \nnb \\
&&+16 C_{TE} \bar \ell \sigma_{\mu\nu}  \ell \Big[ -2 T_1
{\varepsilon^\ast}^\mu (p_B+p_{\rho})^\nu  +B_6 {\varepsilon^\ast}^\mu q^\nu -
B_7 (\varepsilon^\ast q) {p_{\rho}}^\mu q^\nu
\Bigg\}~,
\eea
where
\bea
\label{e7013}
A_1 &=& (C_{LL}^{tot} + C_{RL}) \frac{V}{m_B+m_{\rho}} -
2 (C_{BR}+C_{SL}) \frac{T_1}{q^2} ~, \nnb \\
B_1 &=& (C_{LL}^{tot} - C_{RL}) (m_B+m_{\rho}) A_1 - 2
(C_{BR}-C_{SL}) (m_B^2-m_{\rho}^2)
\frac{T_2}{q^2} ~, \nnb \\
B_2 &=& \frac{C_{LL}^{tot} - C_{RL}}{m_B+m_{\rho}} A_2 - 2
(C_{BR}-C_{SL})
\frac{1}{q^2}  \left[ T_2 + \frac{q^2}{m_B^2-m_{\rho}^2} T_3 \right]~,
\nnb \\
B_3 &=& 2 (C_{LL}^{tot} - C_{RL}) m_{\rho} \frac{A_3-A_0}{q^2}+
2 (C_{BR}-C_{SL}) \frac{T_3}{q^2} ~, \nnb \\
C_1 &=& A_1 ( C_{LL}^{tot} \rar C_{LR}^{tot}~,~~C_{RL} \rar
C_{RR})~,\nnb \\
D_1 &=& B_1 ( C_{LL}^{tot} \rar C_{LR}^{tot}~,~~C_{RL} \rar
C_{RR})~,\nnb \\
D_2 &=& B_2 ( C_{LL}^{tot} \rar C_{LR}^{tot}~,~~C_{RL} \rar
C_{RR})~,\nnb \\
D_3 &=& B_3 ( C_{LL}^{tot} \rar C_{LR}^{tot}~,~~C_{RL} \rar
C_{RR})~,\nnb \\
B_4 &=& - 2 ( C_{LRRL} - C_{RLRL}) \frac{ m_{\rho}}{m_b} A_0 ~,\nnb \\
B_5 &=& - 2 ( C_{LRLR} - C_{RLLR}) \frac{m_{\rho}}{m_b} A_0 ~,\nnb \\
B_6 &=& 2 (m_B^2-m_{\rho}^2) \frac{T_1-T_2}{q^2} ~,\nnb \\
B_7 &=& \frac{4}{q^2} \left( T_1-T_2 - 
\frac{q^2}{m_B^2-m_{\rho}^2} T_3 \right)~.   
\eea

From this expression of the decay amplitude, for the unpolarized differential
decay width we get the following result:
\bea
\label{e7014}
\frac{d\Gamma}{d\hat{s}}(B \rar \rho \ell^+ \ell^-) =
\frac{G^2 \alpha^2 m_B}{2^{14} \pi^5}
\vel V_{tb}V_{td}^\ast \ver^2 \lambda^{1/2}(1,\hat{r},\hat{s}) v
\Delta(\hat{s})~,
\eea
with
\bea
\label{e7015}
\Delta \es
\frac{2 m_B^2}{3 \hat{r}_{\rho} \hat{s}}
\,\mbox{\rm Re}\bigg\{
%1
- 6 m_B \hat{m}_\ell \hat{s} \lambda
(B_1-D_1) (B_4^\ast - B_5^\ast) \nnb \\
%2
\ek 12 m_B^2 \hat{m}_\ell^2 \hat{s} \lambda
\Big[ B_4 B_5^\ast + (B_3-D_2-D_3) B_1^\ast - (B_2+B_3-D_3)
D_1^\ast \Big] \nnb \\
%3
\ar 6 m_B^3 \hat{m}_\ell \hat{s}
(1-\hat{r}_{\rho}) \lambda
(B_2-D_2) (B_4^\ast - B_5^\ast) \nnb \\
%4
\ar 12 m_B^4 \hat{m}_\ell^2 \hat{s} 
(1-\hat{r}_{\rho}) \lambda
(B_2-D_2) (B_3^\ast-D_3^\ast) \nnb \\
%5
\ar 6 m_B^3 \hat{m}_\ell \lambda \hat{s}^2
(B_4-B_5) (B_3^\ast-D_3^\ast) \nnb \\
%6
\ar 48 \hat{m}_\ell^2 \hat{r}_{\rho} \hat{s} \Big( 3 B_1 D_1^\ast +
2 m_B^4 \lambda A_1 C_1^\ast \Big) \nnb \\
%7
\ar 48 m_B^5 \hat{m}_\ell \hat{s}\lambda^2
(B_2+D_2) B_7^\ast C_{TE}^\ast \nnb \\
%8
\ek 16 m_B^4 \hat{r}_{\rho} \hat{s} (\hat{m}_\ell^2-\hat{s}) \lambda
\Big( \vel A_1\ver^2 + \vel C_1\ver^2 \Big) \nnb \\
%9
\ek m_B^2 \hat{s} (2 \hat{m}_\ell^2-\hat{s}) \lambda
\Big( \vel B_4\ver^2 + \vel B_5\ver^2 \Big) \nnb \\
%10
\ek 48 
m_B^3 \hat{m}_\ell \hat{s} (1-\hat{r}_{\rho}-\hat{s}) \lambda
\Big[(B_1+D_1) B_7^\ast C_{TE}^\ast +
2 (B_2+D_2) B_6^\ast C_{TE}^\ast \Big] \nnb \\
%11
\ek 6 m_B^4 \hat{m}_\ell^2 \hat{s} \lambda
\Big[ 2 (2+2\hat{r}_{\rho}-\hat{s}) B_2 D_2^\ast -
\hat{s} \vel (B_3-D_3)\ver^2 \Big] \nnb \\
%12
\ar 96
m_B \hat{m}_\ell \hat{s} (\lambda + 12 \hat{r}_{\rho} \hat{s})
(B_1+D_1) B_6^\ast C_{TE}^\ast\nnb \\
%13
\ar 8 
m_B^2 \hat{s}^2 \Big[
v^2 \vel C_T \ver^2 + 4 (3-2 v^2) \vel C_{TE} \ver^2 \Big]
\Big[ 4 (\lambda + 12 \hat{r}_{\rho} \hat{s}) \vel B_6 \ver^2 \nnb \\ 
\ek 4 m_B^2 \lambda (1-\hat{r}_{\rho}-\hat{s}) B_6 B_7^\ast
+ m_B^4 \lambda^2 \vel B_7 \ver^2  \Big] \nnb \\
%14
\ek 4 m_B^2 \lambda \Big[
\hat{m}_\ell^2 (2 - 2 \hat{r}_{\rho} + \hat{s} ) +
\hat{s} (1 - \hat{r}_{\rho} - \hat{s} ) \Big]
(B_1 B_2^\ast + D_1 D_2^\ast) \nnb \\
%15
\ar \hat{s} \Big[
6 \hat{r}_{\rho} \hat{s} (3+v^2) + \lambda (3-v^2)
\Big] \Big( \vel B_1\ver^2 + \vel D_1\ver^2 \Big) \nnb \\
%16
\ek 2 m_B^4 \lambda \Big\{
\hat{m}_\ell^2 [\lambda - 3 (1-\hat{r}_{\rho})^2] - \lambda \hat{s} \Big\}
\Big( \vel B_2\ver^2 + \vel D_2\ver^2 \Big) \nnb \\
%17
\ar 128 m_B^2 \Big\{
4 \hat{m}_\ell^2 [ 20 \hat{r}_{\rho} \lambda - 
12 \hat{r}_{\rho} (1-\hat{r}_{\rho})^2 - \lambda \hat{s}] \nnb \\ 
\ar \hat{s} [ 4 \hat{r}_{\rho} \lambda + 12 \hat{r}_{\rho} (1-\hat{r}_{\rho})^2 +
\lambda \hat{s}] \Big\}
\vel C_T\ver^2 \vel t_1\ver^2 \nnb \\
%18
\ar 512 m_B^2 \Big\{
\hat{s} [ 4 \hat{r}_{\rho} \lambda + 12 \hat{r}_{\rho} (1-\hat{r}_{\rho})^2 +
\lambda \hat{s}] \nnb \\ 
\ar 8 \hat{m}_\ell^2 [ 12 \hat{r}_{\rho} (1-\hat{r}_{\rho})^2 +
\lambda (\hat{s}-8 \hat{r}_{\rho})] \Big\}
\vel C_{TE}\ver^2 \vel t_1\ver^2 \nnb \\
%19
\ek 64 m_B^2 \hat{s}^2 
\Big[ v^2 \vel C_T \ver^2
+ 4 (3 - 2 v^2) \vel C_{TE} \ver^2 \Big]
\Big\{ 2 [ \lambda  + 12 \hat{r}_{\rho} (1-\hat{r}_{\rho})]
B_6 t_1^\ast \nnb \\ 
\ek m_B^2 \lambda (1 + 3 \hat{r}_{\rho} - \hat{s}) 
B_7 t_1^\ast \Big\} \nnb \\
%20
\ar 768  m_B^3 \hat{m}_\ell \hat{r}_{\rho} \hat{s}
\lambda (A_1 + C_1) C_T^\ast t_1^\ast \nnb \\
%21
\ek 192 m_B \hat{m}_\ell \hat{s}
[ \lambda  + 12 \hat{r}_{\rho} (1-\hat{r}_{\rho})]
(B_1 + D_1) C_{TE}^\ast t_1^\ast \nnb \\
%22
\ar 192 m_B^3 \hat{m}_\ell \hat{s} \lambda
(1 + 3 \hat{r}_{\rho} -\hat{s}) \lambda
(B_2 + D_2) C_{TE}^\ast t_1^\ast 
\bigg\}~,
\eea
where $\hat{s}=q^2/m_B^2$, $\hat{r}_{\rho}=m_{\rho}^2/m_B^2$ and
$\lambda(a,b,c)=a^2+b^2+c^2-2ab-2ac-2bc$,
$\hat{m}_\ell=m_\ell/m_B$, $v=\sqrt{1-4\hat{m}_\ell^2/\hat{s}}$ is the
final lepton velocity.

Using the matrix element for the $B \rar \rho \ell^+ \ell^-$ decay, our next
problem is to calculate the nine double--lepton polarization asymmetries. 
For this aim we introduce the spin projection operators
\bea
\Lambda_1 \es \frac{1}{2} (1+\gamma_5 \not\!{s}_i^- )~,\nnb \\
\Lambda_2 \es \frac{1}{2} (1+\gamma_5 \not\!{s}_i^+ )~\nnb 
\eea
for the lepton $\ell^-$ and anti--lepton $\ell^+$, where
$i=L,N,T$ correspond to the longitudinal, normal and transversal
polarizations, respectively. 
Firstly we define the following orthogonal unit vectors $s_i^{\pm\mu}$ 
in the rest frame of $\ell^\pm$ (see also \cite{R7001,R7003,R7022,R7023}),
\bea
\label{e7016}   
s^{-\mu}_L \es \ga 0,\vec{e}_L^{\,-}\dr =
\ga 0,\frac{\vec{p}_-}{\vel\vec{p}_- \ver}\dr~, \nnb \\
s^{-\mu}_N \es \ga 0,\vec{e}_N^{\,-}\dr = \ga 0,\frac{\vec{p}_\rho\times
\vec{p}_-}{\vel \vec{p}_\rho\times \vec{p}_- \ver}\dr~, \nnb \\
s^{-\mu}_T \es \ga 0,\vec{e}_T^{\,-}\dr = \ga 0,\vec{e}_N^{\,-}
\times \vec{e}_L^{\,-} \dr~, \nnb \\
s^{+\mu}_L \es \ga 0,\vec{e}_L^{\,+}\dr =
\ga 0,\frac{\vec{p}_+}{\vel\vec{p}_+ \ver}\dr~, \nnb \\
s^{+\mu}_N \es \ga 0,\vec{e}_N^{\,+}\dr = \ga 0,\frac{\vec{p}_\rho\times
\vec{p}_+}{\vel \vec{p}_\rho\times \vec{p}_+ \ver}\dr~, \nnb \\
s^{+\mu}_T \es \ga 0,\vec{e}_T^{\,+}\dr = \ga 0,\vec{e}_N^{\,+}
\times \vec{e}_L^{\,+}\dr~,
\eea
where $\vec{p}_\mp$ and $\vec{p}_\rho$ are the three--momenta of the
leptons $\ell^\mp$ and $\rho$ meson in the
center of mass frame (CM) of $\ell^- \,\ell^+$ system, respectively.
Transformation of unit vectors from the rest frame of the leptons to CM
frame of leptons can be done by the Lorentz boost. Boosting of the
longitudinal unit vectors $s_L^{\pm\mu}$ leads to
\bea
\label{e7017}
\ga s^{\mp\mu}_L \dr_{CM} \es \ga \frac{\vel\vec{p}_\mp \ver}{m_\ell}~,
\frac{E_\ell \vec{p}_\mp}{m_\ell \vel\vec{p}_\mp \ver}\dr~,
\eea
where $\vec{p}_+ = - \vec{p}_-$, $E_\ell$ and $m_\ell$ are the energy and mass
of leptons in the CM frame, respectively.
The remaining two unit vectors $s_N^{\pm\mu}$, $s_T^{\pm\mu}$ are unchanged
under Lorentz boost.

We can now define the double--lepton polarization asymmetries as in
\cite{R7008}:
\bea
\label{e7018}  
P_{ij}(\hat{s}) =
\frac{\ds{\Bigg( \frac{d\Gamma}{d\hat{s}}(\vec{s}_i^-,\vec{s}_j^+)}-
\ds{\frac{d\Gamma}{d\hat{s}}(-\vec{s}_i^-,\vec{s}_j^+) \Bigg)} -
\ds{\Bigg( \frac{d\Gamma}{d\hat{s}}(\vec{s}_i^-,-\vec{s}_j^+)} -
\ds{\frac{d\Gamma}{d\hat{s}}(-\vec{s}_i^-,-\vec{s}_j^+)\Bigg)}}
{\ds{\Bigg( \frac{d\Gamma}{d\hat{s}}(\vec{s}_i^-,\vec{s}_j^+)} +
\ds{\frac{d\Gamma}{d\hat{s}}(-\vec{s}_i^-,\vec{s}_j^+) \Bigg)} +
\ds{\Bigg( \frac{d\Gamma}{d\hat{s}}(\vec{s}_i^-,-\vec{s}_j^+)} +
\ds{\frac{d\Gamma}{d\hat{s}}(-\vec{s}_i^-,-\vec{s}_j^+)\Bigg)}}~,
\eea
where $i,j=L,~N,~T$, and the first subindex $i$ corresponds
lepton while the second subindex $j$ corresponds to antilepton,
respectively.

After lengthy calculations we get the following results for the
double--polarization asymmetries.

\bea
\label{e7019}
P_{LL} \es \frac{m_B^2}{3 \hat{r}_{\rho} \hat{s} \Delta}
\, \mbox{\rm Re} \bigg\{
%1
- 12 m_B \hat{m}_\ell \hat{s} \lambda
(B_1-D_1) (B_4^\ast - B_5^\ast) \nnb \\
%2
\ek 24 m_B^2 \hat{m}_\ell^2 \hat{s} \lambda
\Big[ B_4 B_5^\ast + (B_1-D_1) (B_3^\ast - D_3^\ast) \Big] \nnb \\
%3
\ar 12 m_B^3 \hat{m}_\ell \hat{s} \lambda (1-\hat{r}_{\rho})
\Big[ (B_2 - D_2) (B_4^\ast - B_5^\ast) + 2 m_B \hat{m}_\ell
(B_2 - D_2) (B_3^\ast - D_3^\ast) \Big] \nnb \\
%4
\ek 32 m_B^5 \hat{m}_\ell \hat{s} \lambda^2
(B_2 + D_2) B_7^\ast C_{TE}^\ast \nnb \\
%5
\ar 3 m_B^2 \hat{s}^2 \lambda (1+v^2)
(\vel B_4 \ver^2 + \vel B_5 \ver^2) \nnb \\
%6
\ek 8 m_B^4 \hat{r}_{\rho} \hat{s}^2 \lambda (1+3 v^2)
(\vel A_1 \ver^2 + \vel C_1 \ver^2) \nnb \\
%7
\ar 12 m_B^3 \hat{m}_\ell \hat{s}^2 \lambda
(B_3 - D_3) (B_4^\ast - B_5^\ast) \nnb \\
%8
\ar 12 m_B^4 \hat{m}_\ell^2 \hat{s}^2 \lambda
\vel B_3 - D_3 \ver^2 \nnb \\
%9
\ar 32 m_B^3 \hat{m}_\ell \hat{s} \lambda
(1-\hat{r}_{\rho}-\hat{s})
\Big[ (B_1 + D_1) B_7^\ast C_{TE}^\ast +
2 (B_2 + D_2) B_6^\ast C_{TE}^\ast \Big] \nnb \\
%10
\ar 8 m_B^2 \hat{m}_\ell^2 \lambda
(4 - 4 \hat{r}_{\rho} - \hat{s})
(B_1 D_2^\ast + B_2 D_1^\ast) \nnb \\
%11
\ek 64 m_B \hat{m}_\ell \hat{s}
(\lambda + 12 \hat{r}_{\rho} \hat{s})
(B_1 + D_1) B_6^\ast C_{TE}^\ast \nnb \\
%12
\ek 16 m_B^2 \hat{s} \Big[4 \hat{m}_\ell^2
(\vel C_T \ver^2 + 8 \vel C_{TE} \ver^2 ) -
\hat{s} (\vel C_T \ver^2 + 4 \vel C_{TE} \ver^2 )\Big]
\Big[ m_B^4 \lambda^2 \vel B_7 \ver^2 +
4 (\lambda + 12 \hat{r}_{\rho} \hat{s}) \vel B_6 \ver^2 \Big] \nnb \\
%13
\ek 32 \hat{m}_\ell^2
(\lambda + 3 \hat{r}_{\rho} \hat{s}) B_1 D_1^\ast \nnb \\
%14
\ek 8 m_B^4 \hat{m}_\ell^2 \lambda
[\lambda + 3 (1-\hat{r}_{\rho})^2] B_2 D_2^\ast \nnb \\
%15
\ar 8 m_B^2 \lambda
[\hat{s} - \hat{s} (\hat{r}_{\rho} + \hat{s}) - 3 \hat{m}_\ell^2
(2 - 2 \hat{r}_{\rho} - \hat{s})] (B_1 B_2^\ast + D_1 D_2^\ast) \nnb \\
%16
\ek 64 m_B^4 \hat{s}^2 \lambda
(1 - \hat{r}_{\rho} - \hat{s})
\Big[ v^2 \vel C_T \ver^2 - 4 (1-2 v^2) \vel C_{TE} \ver^2 \Big]
B_6 B_7^\ast\nnb \\
%17
\ek m_B^4 \hat{s} \lambda
[\lambda (1+3 v^2) - 3 (1-\hat{r}_{\rho})^2 (1-v^2)]
(\vel B_2 \ver^2 + \vel D_2 \ver^2) \nnb \\
%18
\ar 4 [6 \hat{m}_\ell^2 (\lambda + 6 \hat{r}_{\rho}\hat{s})
- \hat{s} (\lambda + 12 \hat{r}_{\rho}\hat{s})]
(\vel B_1 \ver^2 + \vel D_1 \ver^2) \nnb \\
%19
\ek 1024 m_B^2
\{12 \hat{r}_{\rho} \hat{s} (1-\hat{r}_{\rho})^2 (1-2 v^2)
- \lambda \hat{s} [ 4 \hat{r}_{\rho} - \hat{s} (1-2 v^2)] \}
\vel t_1 \ver^2 \vel C_{TE} \ver^2 \nnb \\
%20
\ar 256 m_B^2 \hat{s}
\{\lambda (\hat{s} v^2 - 8 \hat{r}_{\rho}) +
12 \hat{r}_{\rho} v^2 [\lambda + (1-\hat{r}_{\rho})^2] \}
\vel t_1 \ver^2 \vel C_T \ver^2 \nnb \\
%21
\ek 256 m_B^2 \hat{s}^2
[\lambda + 12 \hat{r}_{\rho} (1-\hat{r}_{\rho})]
\Big[ v^2 \vel C_T \ver^2 - 4 (1-2 v^2) \vel C_{TE} \ver^2 \Big]
B_6 t_1^\ast \nnb \\
%22
\ar 128 m_B^4 \hat{s}^2 \lambda
(1+ 3 \hat{r}_{\rho} - \hat{s})
\Big[ v^2 \vel C_T \ver^2 - 4 (1-2 v^2) \vel C_{TE} \ver^2 \Big]
B_7 t_1^\ast \nnb \\
%23
\ar 128 m_B \hat{m}_\ell \hat{s}
[\lambda + 12 \hat{r}_{\rho} (1-\hat{r}_{\rho})]
(B_1 + D_1) t_1^\ast C_{TE}^\ast \nnb \\
%24
\ek 128 m_B^3 \hat{m}_\ell \hat{s} \lambda
(1+ 3 \hat{r}_{\rho} - \hat{s})
(B_2 + D_2) t_1^\ast C_{TE}^\ast \nnb \\
%25
\ek 512 m_B^3 \hat{m}_\ell \hat{r}_{\rho} \hat{s} \lambda
(A_1 + C_1) t_1^\ast C_T^\ast \nnb \\
%26
\ek 64 m_B^4 \hat{m}_\ell^2 \hat{r}_{\rho} \hat{s} \lambda
A_1 C_1^\ast \bigg\}
~, \\ \nnb \\
\label{e7020}
P_{LN} \es
\frac{\pi m_B^2}{2\hat{r}_{\rho}\Delta} \sqrt{\frac{\lambda}{\hat{s}}}
\, \mbox{\rm Im} \bigg\{
%1
4 m_B^2 \hat{m}_\ell^2 \lambda
\Big[B_2 B_4^\ast + B_5 D_2^\ast +
8 (B_1 - D_1) B_7^\ast C_{TE}^\ast \Big] \nnb \\
%2
\ek 4 m_B^4 \hat{m}_\ell \lambda (1-\hat{r}_{\rho})
\Big[B_2 D_2^\ast +
8 m_B \hat{m}_\ell (B_2 - D_2) B_7^\ast C_{TE}^\ast \Big] \nnb \\
%3
\ar 2 m_B^4 \hat{m}_\ell \hat{s} \lambda
\Big[B_2 B_3^\ast - 8 (B_4 - B_5) B_7^\ast C_{TE}^\ast
- 16 m_B \hat{m}_\ell (B_3 - D_3) B_7^\ast C_{TE}^\ast
\Big] \nnb \\
%4
\ek 2 m_B^4 \hat{m}_\ell \hat{s} \lambda
\Big[ B_3 D_2^\ast + (B_2 + D_2) D_3^\ast \Big] \nnb \\
%5
\ek 2 m_B^2 \hat{m}_\ell \hat{s} (1+3 \hat{r}_{\rho} -\hat{s})
\Big( B_1 B_2^\ast - D_1 D_2^\ast - 
32 B_5 C_{TE}^\ast t_1^\ast \Big) \nnb \\
%6
\ar 32 m_B^3 \hat{r}_{\rho} \hat{s}^2 v^2
\Big[ (A_1 - C_1) B_6^\ast C_T^\ast -
2 (A_1 + C_1) B_6^\ast C_{TE}^\ast \Big] \nnb \\
%7
\ek m_B^3 \hat{s} \lambda
(1+v^2) \Big( B_2 B_5^\ast + B_4 D_2^\ast \Big) \nnb \\
%8
\ek 4 \hat{m}_\ell (1-\hat{r}_{\rho}-\hat{s})
\Big\{ B_1 D_1^\ast + m_B \hat{m}_\ell \Big[
B_1 (B_4^\ast + 16 B_6^\ast C_{TE}^\ast) -
D_1 (B_5^\ast + 16 B_6^\ast C_{TE}^\ast) \Big] \Big\} \nnb \\
%9
\ar 64 m_B^3 \hat{m}_\ell^2
(1-\hat{r}_{\rho}) (1-\hat{r}_{\rho}-\hat{s})
(B_2 - D_2) B_6^\ast C_{TE}^\ast \nnb \\
%10
\ek 2 m_B^2 \hat{m}_\ell \hat{s} (1-\hat{r}_{\rho}-\hat{s})
(B_1 + D_1) (B_3^\ast - D_3^\ast) \nnb \\
%11
\ar 32 m_B^2 \hat{m}_\ell \hat{s} (1-\hat{r}_{\rho}-\hat{s})
\Big[ (B_4 - B_5) B_6^\ast C_{TE}^\ast
+ 2 m_B \hat{m}_\ell (B_3 - D_3) B_6^\ast C_{TE}^\ast \Big] \nnb \\
%12\
\ar m_B \hat{s} (1-\hat{r}_{\rho}-\hat{s}) (1+v^2)
\Big( B_1 B_5^\ast + B_4 D_1^\ast \Big) \nnb \\
%13
\ar 2 m_B^2 \hat{m}_\ell
[\lambda + (1-\hat{r}_{\rho}) (1-\hat{r}_{\rho}-\hat{s})]
\Big( B_2 D_1^\ast + B_1 D_2^\ast \Big) \nnb \\
%14
\ek 128 m_B^3 \hat{m}_\ell^2
(1-\hat{r}_{\rho}) (1+ 3 \hat{r}_{\rho}-\hat{s})
(B_2 - D_2) t_1^\ast C_{TE}^\ast \nnb \\
%15
\ek 64 m_B^2 \hat{m}_\ell \hat{s}
(1+ 3 \hat{r}_{\rho}-\hat{s})
\Big[B_4 C_{TE}^\ast t_1^\ast +
2 m_B \hat{m}_\ell (B_3 - D_3) C_{TE}^\ast t_1^\ast \Big] \nnb \\
%16
\ek 64 m_B^3 \hat{r}_{\rho} \hat{s} (1- \hat{r}_{\rho}) v^2
\Big[ A_1 (C_T^\ast - 2 C_{TE}^\ast) t_1^\ast -
C_1 (C_T^\ast + 2 C_{TE}^\ast) t_1^\ast \Big] \nnb \\
%17
\ek 32 m_B \hat{s}
\Big[(1+ 3 \hat{r}_{\rho}-\hat{s}) D_1 C_{TE}^\ast t_1^\ast
+ 2 \hat{r}_{\rho} v^2 D_1 C_T^\ast t_1^\ast
- (1-\hat{r}_{\rho} - \hat{s}) v^2 D_1 C_{TE}^\ast t_1^\ast \Big] \nnb \\
%18
\ar 32 m_B \hat{s}
\Big[(1+ 3 \hat{r}_{\rho}-\hat{s}) B_1 C_{TE}^\ast t_1^\ast
- 2 \hat{r}_{\rho} v^2 B_1 C_T^\ast t_1^\ast
- (1-\hat{r}_{\rho} - \hat{s}) v^2 B_1 C_{TE}^\ast t_1^\ast 
\Big]
\bigg\}~, \\ \nnb \\
\label{e7021}
P_{NL} \es
\frac{\pi m_B^2}{2\hat{r}_{\rho}\Delta} \sqrt{\frac{\lambda}{\hat{s}}}
\, \mbox{\rm Im} \bigg\{
%1
4 m_B^3 \hat{m}_\ell^2 \lambda
\Big[B_2 B_5^\ast + B_4 D_2^\ast -
8 (B_1 - D_1) B_7^\ast C_{TE}^\ast \Big] \nnb \\
%2
\ar 4 m_B^4 \hat{m}_\ell \lambda (1-\hat{r}_{\rho})
\Big[B_2 D_2^\ast +
8 m_B \hat{m}_\ell (B_2 - D_2) B_7^\ast C_{TE}^\ast \Big] \nnb \\
%3
\ek 2 m_B^4 \hat{m}_\ell \hat{s} \lambda
\Big[B_2 B_3^\ast - 8 (B_4 - B_5) B_7^\ast C_{TE}^\ast
- 16 m_B \hat{m}_\ell (B_3 - D_3) B_7^\ast C_{TE}^\ast
\Big] \nnb \\
%4
\ar 2 m_B^4 \hat{m}_\ell \hat{s} \lambda
\Big[ B_3 D_2^\ast + (B_2 + D_2) D_3^\ast \Big] \nnb \\
%5
\ar 2 m_B^2 \hat{m}_\ell
\hat{s} (1+3 \hat{r}_{\rho} -\hat{s})
\Big( B_1 B_2^\ast - D_1 D_2^\ast - 
32 B_5 C_{TE}^\ast t_1^\ast \Big) \nnb \\
%6
\ek 32 m_B^3 \hat{r}_{\rho} \hat{s}^2 v^2
\Big[ (A_1 - C_1) B_6^\ast C_T^\ast +
2 (A_1 + C_1) B_6^\ast C_{TE}^\ast \Big] \nnb \\
%7
\ek m_B^3 \hat{s} \lambda
(1+v^2) \Big( B_2 B_4^\ast + B_5 D_2^\ast \Big) \nnb \\
%8
\ar 4 \hat{m}_\ell (1-\hat{r}_{\rho}-\hat{s})
\Big\{ B_1 D_1^\ast - m_B \hat{m}_\ell \Big[
B_1 (B_5^\ast - 16 B_6^\ast C_{TE}^\ast) -
D_1 (B_4^\ast - 16 B_6^\ast C_{TE}^\ast) \Big] \Big\} \nnb \\
%9
\ar 64 m_B^3 \hat{m}_\ell^2
(1-\hat{r}_{\rho}) (1-\hat{r}_{\rho}-\hat{s})
(B_2 - D_2) B_6^\ast C_{TE}^\ast \nnb \\
%10
\ar 2 m_B^2 \hat{m}_\ell
\hat{s} (1-\hat{r}_{\rho}-\hat{s})
(B_1 + D_1) (B_3^\ast - D_3^\ast) \nnb \\
%11
\ek 32 m_B^2 \hat{m}_\ell
\hat{s} (1-\hat{r}_{\rho}-\hat{s})
\Big[ (B_4 - B_5) B_6^\ast C_{TE}^\ast
+ 2 m_B \hat{m}_\ell (B_3 - D_3) B_6^\ast C_{TE}^\ast \Big] \nnb \\
%12
\ar m_B \hat{s} (1-\hat{r}_{\rho}-\hat{s}) (1+v^2)
\Big( B_1 B_4^\ast + B_5 D_1^\ast \Big) \nnb \\
%13
\ek 2 m_B^2 \hat{m}_\ell
[\lambda + (1-\hat{r}_{\rho}) (1-\hat{r}_{\rho}-\hat{s})]
\Big( B_2 D_1^\ast + B_1 D_2^\ast \Big) \nnb \\
%14
\ar 128 m_B^3 \hat{m}_\ell^2
(1-\hat{r}_{\rho}) (1+ 3 \hat{r}_{\rho}-\hat{s})
(B_2 - D_2) t_1^\ast C_{TE}^\ast \nnb \\
%15
\ar 64 m_B^2 \hat{m}_\ell
\hat{s} (1+ 3 \hat{r}_{\rho}-\hat{s})
\Big[B_4 C_{TE}^\ast t_1^\ast +
2 m_B \hat{m}_\ell (B_3 - D_3) C_{TE}^\ast t_1^\ast \Big] \nnb \\
%16
\ar 64 m_B^3\hat{r}_{\rho} \hat{s}
(1- \hat{r}_{\rho}) v^2
\Big[ A_1 (C_T^\ast + 2 C_{TE}^\ast) t_1^\ast -
C_1 (C_T^\ast - 2 C_{TE}^\ast) t_1^\ast \Big] \nnb \\
%17
\ek 32 m_B \hat{s}
\Big[(1+ 3 \hat{r}_{\rho}-\hat{s}) B_1 C_{TE}^\ast t_1^\ast
+ 2 \hat{r}_{\rho} v^2 B_1 C_T^\ast t_1^\ast
- (1-\hat{r}_{\rho} - \hat{s}) v^2 B_1 C_{TE}^\ast t_1^\ast \Big] \nnb \\
%18
\ar 32 m_B \hat{s}
\Big[(1+ 3 \hat{r}_{\rho}-\hat{s}) D_1 C_{TE}^\ast t_1^\ast
- 2 \hat{r}_{\rho} v^2 D_1 C_T^\ast t_1^\ast
- (1-\hat{r}_{\rho} - \hat{s}) v^2 D_1 C_{TE}^\ast t_1^\ast 
\Big]
\bigg\}~, \\ \nnb \\
\label{e7022}
P_{LT} \es
\frac{\pi m_B^2 v}{\hat{r}_{\rho}\Delta} \sqrt{\frac{\lambda}{\hat{s}}}
\, \mbox{\rm Re} \bigg\{
%1
m_B^4 \hat{m}_\ell \lambda (1-\hat{r}_{\rho})
\vel B_2 - D_2 \ver^2 \nnb \\
%2
\ek 8 m_B^2 \hat{m}_\ell \hat{r}_{\rho} \hat{s}
\Big( A_1 B_1^\ast - C_1 D_1^\ast \Big) \nnb \\
%3
\ek m_B^3 \hat{s} \lambda
\Big(B_2 B_5^\ast + B_4 D_2^\ast - 
m_B \hat{m}_\ell B_2 B_3^\ast \Big) \nnb \\
%4
\ek 8 m_B^4 \hat{m}_\ell \hat{s}
\lambda (B_4+B_5) B_7^\ast C_{TE}^\ast \nnb \\
%5
\ek m_B^4 \hat{m}_\ell \hat{s}
\lambda \Big(B_2 D_3^\ast + 
B_3 D_2^\ast - D_2 D_3^\ast \Big) \nnb \\
%6
\ar 16 m_B^3 \hat{r}_{\rho} \hat{s}^2
\Big[ A_1 B_6^\ast (C_T^\ast - 2 C_{TE}^\ast) +
      C_1 B_6^\ast (C_T^\ast + 2 C_{TE}^\ast)\Big] \nnb \\
%7
\ar \hat{m}_\ell (1-\hat{r}_{\rho} -\hat{s})
\vel B_1 - D_1 \ver^2 \nnb \\
%8
\ar m_B \hat{s} (1-\hat{r}_{\rho} -\hat{s})
\Big[ B_1 B_5^\ast + B_4 D_1^\ast +
16 m_B \hat{m}_\ell (B_4 +B_5) B_6^\ast C_{TE}^\ast \nnb \\
\ek m_B \hat{m}_\ell (B_1 - D_1) (B_3^\ast - D_3^\ast) \Big] \nnb \\
%9
\ek m_B^2 \hat{m}_\ell
[\lambda + (1-\hat{r}_{\rho}) (1-\hat{r}_{\rho} -\hat{s})]
(B_1 - D_1) (B_2^\ast - D_2^\ast) \nnb \\
%10
\ek 1024 m_B^2 \hat{m}_\ell \hat{r}_{\rho} (1-\hat{r}_{\rho})
\Big( \vel C_T \ver^2 + 4 \vel C_{TE} \ver^2 \Big)
\vel t_1 \ver^2 \nnb \\
%11
\ar 512 m_B^2 \hat{m}_\ell \hat{r}_{\rho} \hat{s}
\Big( \vel C_T \ver^2 + 4 \vel C_{TE} \ver^2 \Big)
B_6 t_1^\ast \nnb \\
%12
\ek 32 m_B^2 \hat{m}_\ell \hat{s} (1+3\hat{r}_{\rho}-\hat{s})
(B_4 + B_5) C_{TE}^\ast t_1^\ast \nnb \\
%13
\ek 32 m_B^3 \hat{r}_{\rho} \hat{s} (1-\hat{r}_{\rho})
\Big[ A_1 (C_T^\ast - 2 C_{TE}^\ast) t_1^\ast +
C_1 (C_T^\ast + 2 C_{TE}^\ast) t_1^\ast \Big] \nnb \\
%14
\ek 32 m_B \hat{r}_{\rho} \hat{s}
\Big[ B_1 (C_T^\ast - 2 C_{TE}^\ast) t_1^\ast -
D_1 (C_T^\ast + 2 C_{TE}^\ast) t_1^\ast \Big]
\bigg\}~, \\ \nnb \\
\label{e7023}
P_{TL} \es
\frac{\pi m_B^2 v}{\hat{r}_{\rho}\Delta} \sqrt{\frac{\lambda}{\hat{s}}}
\, \mbox{\rm Re} \bigg\{
%1
m_B^4 \hat{m}_\ell \lambda (1-\hat{r}_{\rho})
\vel B_2 - D_2 \ver^2 \nnb \\
%2
\ar 8 m_B^2 \hat{m}_\ell \hat{r}_{\rho} \hat{s}
\Big( A_1 B_1^\ast - C_1 D_1^\ast \Big) \nnb \\
%3
\ar m_B^3 \hat{s} \lambda
\Big(B_2 B_4^\ast + B_5 D_2^\ast + 
m_B \hat{m}_\ell B_2 B_3^\ast \Big) \nnb \\
%4
\ar 8 m_B^4 \hat{m}_\ell \hat{s}
\lambda (B_4+B_5) B_7^\ast C_{TE}^\ast \nnb \\
%5
\ek m_B^4 \hat{m}_\ell \hat{s}
\lambda \Big(B_2 D_3^\ast + 
B_3 D_2^\ast - D_2 D_3^\ast \Big) \nnb \\
%6
\ar 16 m_B^3 \hat{r}_{\rho} \hat{s}^2
\Big[ A_1 B_6^\ast (C_T^\ast + 2 C_{TE}^\ast) +
      C_1 B_6^\ast (C_T^\ast - 2 C_{TE}^\ast)\Big] \nnb \\
%7
\ar \hat{m}_\ell (1-\hat{r}_{\rho} -\hat{s})
\vel B_1 - D_1 \ver^2 \nnb \\
%8
\ek m_B \hat{s} (1-\hat{r}_{\rho} -\hat{s})
\Big[ B_1 B_4^\ast + B_5 D_1^\ast +
16 m_B \hat{m}_\ell (B_4 +B_5) B_6^\ast C_{TE}^\ast \nnb \\
\ar m_B \hat{m}_\ell (B_1 - D_1) (B_3^\ast - D_3^\ast) \Big] \nnb \\
%9
\ek m_B^2 \hat{m}_\ell
[\lambda + (1-\hat{r}_{\rho}) (1-\hat{r}_{\rho} -\hat{s})]
(B_1 - D_1) (B_2^\ast - D_2^\ast) \nnb \\
%10
\ek 1024 m_B^2 \hat{m}_\ell \hat{r}_{\rho} (1-\hat{r}_{\rho})
\Big( \vel C_T \ver^2 + 4 \vel C_{TE} \ver^2 \Big)
\vel t_1 \ver^2 \nnb \\
%11
\ar 512 m_B^2 \hat{m}_\ell \hat{r}_{\rho} \hat{s}
\Big( \vel C_T \ver^2 + 4 \vel C_{TE} \ver^2 \Big)
B_6 t_1^\ast \nnb \\
%12
\ar 32 m_B^2 \hat{m}_\ell \hat{s} (1+3\hat{r}_{\rho}-\hat{s})
(B_4 + B_5) C_{TE}^\ast t_1^\ast \nnb \\
%13
\ek 32 m_B^3 \hat{r}_{\rho} \hat{s} (1-\hat{r}_{\rho})
\Big[ A_1 (C_T^\ast + 2 C_{TE}^\ast) t_1^\ast +
C_1 (C_T^\ast - 2 C_{TE}^\ast) t_1^\ast \Big] \nnb \\
%14
\ar 32 m_B \hat{r}_{\rho} \hat{s}
\Big[ B_1 (C_T^\ast + 2 C_{TE}^\ast) t_1^\ast -
D_1 (C_T^\ast - 2 C_{TE}^\ast) t_1^\ast \Big]
\bigg\}~, \\ \nnb \\
\label{e7024}
P_{NT} \es
\frac{2 m_B^2 v}{3 \hat{r}_{\rho}\Delta}
\, \mbox{\rm Im} \bigg\{
%1
4 \lambda \Big\{B_1 D_1^\ast +
m_B^4 \lambda \Big[B_2 D_2^\ast -
2 m_B \hat{m}_\ell B_2 B_7^\ast (C_T^\ast - 4 C_{TE}^\ast) \nnb \\
\ek 2 m_B \hat{m}_\ell D_2 B_7^\ast (C_T^\ast + 4 C_{TE}^\ast)
\Big] \Big\} \nnb \\
%2
\ek 6 m_B \hat{m}_\ell \lambda
(B_1 - D_1) (B_4^\ast + B_5^\ast) \nnb \\
%3
\ar 6 m_B^3 \hat{m}_\ell \lambda (1-\hat{r}_{\rho})
(B_2 - D_2) (B_4^\ast + B_5^\ast) \nnb \\
%4
\ar 6 m_B^3 \hat{m}_\ell \hat{s} \lambda
(B_3 - D_3) (B_4^\ast + B_5^\ast) \nnb \\
%5
\ek 4 m_B^2 \lambda (1-\hat{r}_{\rho}-\hat{s})
\Big[ B_1 D_2^\ast + B_2 D_1^\ast +
32 m_B^2 \hat{s} \, \mbox{\rm Re}[B_6 B_7^\ast] C_T C_{TE}^\ast \Big] \nnb \\
%6
\ar 8 m_B^3 \hat{m}_\ell \lambda
(1-\hat{r}_{\rho}-\hat{s})
\Big[ (B_1 B_7^\ast + 2 B_2 B_6^\ast) (C_T^\ast - 4 C_{TE}^\ast) \nnb \\
\ar (B_7^\ast D_1 + 2 B_6^\ast D_2) (C_T^\ast + 4 C_{TE}^\ast)\Big] \nnb \\
%7
\ar 32 m_B^2 \hat{s}
\Big[ 4 (\lambda + 12 \hat{r}_{\rho}\hat{s}) \vel B_6 \ver^2
+ \lambda^2 m_B^4 \vel B_7 \ver^2 \Big]
C_T C_{TE}^\ast \nnb \\
%8
\ar 2 m_B^2 \hat{s} \lambda
\Big( 3 B_4 B_5^\ast - 8 m_B^2 \hat{r}_{\rho} A_1 C_1^\ast \Big) \nnb \\
%9
\ek 16 m_B \hat{m}_\ell
\Big\{ \lambda \Big[ B_1 B_6^\ast (C_T^\ast - 4 C_{TE}^\ast)
+D_1 B_6^\ast (C_T^\ast + 4 C_{TE}^\ast)\Big]
+ 12 \hat{r}_{\rho} \hat{s} (B_1 + D_1) B_6^\ast C_T^\ast \Big\} \nnb \\
%10
\ar 32 m_B \hat{m}_\ell
\Big\{12 \hat{r}_{\rho} (1-\hat{r}_{\rho}) (B_1+D_1) C_T^\ast t_1^\ast
+ \lambda \Big[ B_1 (C_T^\ast - 4 C_{TE}^\ast) t_1^\ast
+ D_1 (C_T^\ast + 4 C_{TE}^\ast) t_1^\ast \Big] \Big\} \nnb \\
%11
\ek 256 m_B^3 \hat{m}_\ell \hat{r}_{\rho} \lambda
\Big[ A_1 (C_T^\ast + 2 C_{TE}^\ast) t_1^\ast - 
C_1 (C_T^\ast - 2 C_{TE}^\ast) t_1^\ast \Big] \nnb \\
%12
\ek 32 m_B^3 \hat{m}_\ell \lambda
(1+3 \hat{r}_{\rho} -\hat{s})
\Big[ B_2 (C_T^\ast - 4 C_{TE}^\ast) t_1^\ast
+D_2 (C_T^\ast + 4 C_{TE}^\ast) t_1^\ast \Big] \nnb \\
%13
\ar 256 m_B^2 \hat{s}
\Big\{ 2 [\lambda + 12 \hat{r}_{\rho} (2 + 2 \hat{r}_{\rho} -\hat{s})]
\vel t_1 \ver^2 -
2 [ \lambda + 12 \hat{r}_{\rho} (1 - \hat{r}_{\rho})]
\mbox{\rm Re} [B_6 t_1^\ast] \nnb \\
\ar m_B^2 \lambda (1+3 \hat{r}_{\rho} -\hat{s})
\mbox{\rm Re} [B_7 t_1^\ast] \Big\} C_T C_{TE}^\ast
\bigg\}~, \\ \nnb \\
\label{e7025}
P_{TN} \es
\frac{2 m_B^2 v}{3 \hat{r}_{\rho}\Delta}
\, \mbox{\rm Im} \bigg\{
%1
- 4 \lambda
\Big\{B_1 D_1^\ast +
m_B^4 \lambda \Big[B_2 D_2^\ast +
2 m_B \hat{m}_\ell B_2 B_7^\ast (C_T^\ast + 4 C_{TE}^\ast) \nnb \\
\ar 2 m_B \hat{m}_\ell D_2 B_7^\ast (C_T^\ast - 4 C_{TE}^\ast)
\Big] \Big\} \nnb \\
%2
\ek 6 m_B \hat{m}_\ell \lambda
(B_1 - D_1) (B_4^\ast + B_5^\ast) \nnb \\
%3
\ar 6 m_B^3 \hat{m}_\ell \lambda (1-\hat{r}_{\rho})
(B_2 - D_2) (B_4^\ast + B_5^\ast) \nnb \\
%4
\ar 6 m_B^3 \hat{m}_\ell \hat{s} \lambda
(B_3 - D_3) (B_4^\ast + B_5^\ast) \nnb \\
%5
\ar 4 m_B^2 \lambda (1-\hat{r}_{\rho}-\hat{s})
\Big[ B_1 D_2^\ast + B_2 D_1^\ast -
32 m_B^2 \hat{s} \, \mbox{\rm Re}[B_6 B_7^\ast] C_T C_{TE}^\ast \Big] \nnb \\
%6
\ar 8 m_B^3 \hat{m}_\ell \lambda
(1-\hat{r}_{\rho}-\hat{s})
\Big[ (B_1 B_7^\ast + 2 B_2 B_6^\ast) (C_T^\ast + 4 C_{TE}^\ast) \nnb \\
\ar (B_7^\ast D_1 + 2 B_6^\ast D_2) (C_T^\ast - 4 C_{TE}^\ast)\Big] \nnb \\
%7
\ar 32 m_B^2 \hat{s}
\Big[ 4 (\lambda + 12 \hat{r}_{\rho}\hat{s}) \vel B_6 \ver^2
+ \lambda^2 m_B^4 \vel B_7 \ver^2 \Big]
C_T C_{TE}^\ast \nnb \\
%8
\ar 2 m_B^2 \hat{s} \lambda
\Big( 3 B_4 B_5^\ast + 8 m_B^2 \hat{r}_{\rho} A_1 C_1^\ast \Big) \nnb \\
%9
\ek 16 m_B \hat{m}_\ell
\Big\{ \lambda \Big[ B_1 B_6^\ast (C_T^\ast + 4 C_{TE}^\ast)
+D_1 B_6^\ast (C_T^\ast - 4 C_{TE}^\ast)\Big]
+ 12 \hat{r}_{\rho} \hat{s} (B_1 + D_1) B_6^\ast C_T^\ast \Big\} \nnb \\
%10
\ar 32 m_B \hat{m}_\ell
\Big\{12 \hat{r}_{\rho} (1-\hat{r}_{\rho}) (B_1+D_1) C_T^\ast t_1^\ast
+ \lambda \Big[ B_1 (C_T^\ast + 4 C_{TE}^\ast) t_1^\ast
+ D_1 (C_T^\ast - 4 C_{TE}^\ast) t_1^\ast \Big] \Big\} \nnb \\
%11
\ar 256 m_B^3 \hat{m}_\ell \hat{r}_{\rho} \lambda
\Big[ A_1 (C_T^\ast - 2 C_{TE}^\ast) t_1^\ast -
C_1 (C_T^\ast + 2 C_{TE}^\ast) t_1^\ast \Big] \nnb \\
%12
\ek 32 m_B^3 \hat{m}_\ell \lambda
(1+3 \hat{r}_{\rho} -\hat{s})
\Big[ B_2 (C_T^\ast + 4 C_{TE}^\ast) t_1^\ast
+D_2 (C_T^\ast - 4 C_{TE}^\ast) t_1^\ast \Big] \nnb \\
%13
\ar 256 m_B^2 \hat{s}
\Big\{ 2 [\lambda + 12 \hat{r}_{\rho} (2 + 2 \hat{r}_{\rho} -\hat{s})]
\vel t_1 \ver^2 -
2 [ \lambda + 12 \hat{r}_{\rho} (1 - \hat{r}_{\rho})]
\mbox{\rm Re} [B_6 t_1^\ast] \nnb \\
\ar m_B^2 \lambda (1+3 \hat{r}_{\rho} -\hat{s})
\mbox{\rm Re} [B_7 t_1^\ast] \Big\} C_T C_{TE}^\ast
\bigg\}~, \\ \nnb \\
\label{e7026}
P_{NN} \es
\frac{2 m_B^2}{3 \hat{r}_{\rho}\Delta}
\, \mbox{\rm Re} \bigg\{
%1
- 24 \hat{m}_\ell^2 \hat{r}_{\rho}
(\vel B_1 \ver^2 + \vel D_1 \ver^2) \nnb \\
%2
\ek 6 m_B \hat{m}_\ell \lambda
(B_1-D_1) (B_4^\ast - B_5^\ast) \nnb \\
%3
\ek 48 m_B^5 \hat{m}_\ell \lambda^2
(B_2 + D_2) B_7^\ast C_{TE}^\ast \nnb \\
%4
\ar 6 m_B^2 \hat{m}_\ell^2 \lambda
\Big[ \vel B_4 \ver^2 +  \vel B_5 \ver^2 -
2 B_1 (B_2^\ast + B_3^\ast - D_3^\ast) +
2 D_1 (B_3^\ast - D_2^\ast - D_3^\ast) \Big] \nnb \\
%5
\ar 6 m_B^3 \hat{m}_\ell \lambda (1-\hat{r}_{\rho})
\Big[ (B_2 - D_2) (B_4^\ast - B_5^\ast) + 2 m_B \hat{m}_\ell
(B_2 - D_2) (B_3^\ast - D_3^\ast) \Big] \nnb \\
%6
\ar m_B^2 \hat{s} \lambda
\Big[ 16 m_B^2 \hat{r}_{\rho} v^2  A_1 C_1^\ast -
3 (1+v^2) B_4 B_5^\ast \big] \nnb \\
%7
\ar 6 m_B^4 \hat{m}_\ell^2 \lambda
(2+2 \hat{r}_{\rho}-\hat{s})
(\vel B_2 \ver^2 + \vel D_2 \ver^2) \nnb \\
%8
\ar 6  m_B^3 \hat{m}_\ell \hat{s} \lambda
(B_3-D_3) (B_4^\ast - B_5^\ast) \nnb \\
%9
\ar 6 m_B^4 \hat{m}_\ell^2 \hat{s} \lambda
\vel B_3 - D_3 \ver^2 \nnb \\
%10
\ar 48 m_B^3 \hat{m}_\ell \lambda (1-\hat{r}_{\rho}-\hat{s})
\Big[ (B_1 + D_1) B_7^\ast C_{TE}^\ast +
2 (B_2 + D_2) B_6^\ast C_{TE}^\ast \Big] \nnb \\
%11
\ek 96 m_B \hat{m}_\ell
(1-\hat{r}_{\rho}-\hat{s})^2
(B_1 + D_1) B_6^\ast C_{TE}^\ast \nnb \\
%12
\ar 8 m_B^4 \hat{s} \lambda
\Big[ \lambda m_B^2 \vel B_7 \ver^2 -
4 (1-\hat{r}_{\rho} - \hat{s}) B_6 B_7^\ast \Big]
\Big[ v^2 \vel C_T \ver^2 - 4 (3-2 v^2) \vel C_{TE} \ver^2 \Big] \nnb \\
%13
\ar m_B^2 \lambda
[3 (2 - 2 \hat{r}_{\rho} - \hat{s}) - v^2 (2 - 2 \hat{r}_{\rho} + \hat{s})]
(B_1 D_2^\ast + B_2 D_1^\ast) \nnb \\
%14
\ek m_B^4 \lambda
\Big[ (3+v^2) \lambda + 3 (1-v^2) (1-\hat{r}_{\rho})^2 \Big]
B_2 D_2^\ast \nnb \\
%15
\ek 2 [6 \hat{r}_{\rho} \hat{s} (1-v^2) + \lambda (3-v^2)]
B_1 D_1^\ast \nnb \\
%16
\ar 32 m_B^2 \hat{s}
\Big\{ (\lambda + 12 \hat{r}_{\rho} \hat{s}) v^2 \vel C_T \ver^2 -
4 [ \lambda (3-2 v^2) +  12 \hat{r}_{\rho} \hat{s}]
\vel C_{TE} \ver^2 \Big\} \vel B_6 \ver^2 \nnb \\
%17
\ek 192 m_B^3 \hat{m}_\ell \lambda
(1+ 3 \hat{r}_{\rho} - \hat{s})
(B_2 + D_2) C_{TE}^\ast t_1^\ast \nnb \\
%18
\ar 192 m_B \hat{m}_\ell
[\lambda + 4 \hat{r}_{\rho} (1-\hat{r}_{\rho})]
(B_1 + D_1) C_{TE}^\ast t_1^\ast \nnb \\
%19
\ar 128 m_B^2 \hat{s} v^2
[\lambda + 12 \hat{r}_{\rho} (2+2 \hat{r}_{\rho}-\hat{s})]
\vel C_T \ver^2 \vel t_1 \ver^2 \nnb \\
%20
\ek 512 m_B^2 \hat{s}
[ \lambda (3 -2 v^2) + 12 \hat{r}_{\rho} (2+2 \hat{r}_{\rho} -\hat{s})]
\vel C_{TE} \ver^2 \vel t_1 \ver^2 \nnb \\
%21
\ek 128 m_B^2 \hat{s}
\Big\{ [\lambda + 12 \hat{r}_{\rho} (1-\hat{r}_{\rho})]
v^2 \vel C_T \ver^2 -
4 [\lambda(3-2 v^2) + 12 \hat{r}_{\rho} (1-\hat{r}_{\rho})]
\vel C_{TE} \ver^2 \Big\}
B_6 t_1^\ast \nnb \\
%22
\ar 64 m_B^4
(1+ 3 \hat{r}_{\rho} - \hat{s}) \hat{s}
\Big[ v^2 \vel C_T \ver^2 - 4 (3-2 v^2) \vel C_{TE} \ver^2 \Big]
B_7 t_1^\ast
\bigg\}~, \\ \nnb \\
\label{e7027}
P_{TT} \es
\frac{2 m_B^2}{3 \hat{r}_{\rho} \hat{s}\Delta}
\, \mbox{\rm Re} \bigg\{
%1
8 m_B^4 \hat{r}_{\rho} \hat{s} \lambda
\Big[ 4 \hat{m}_\ell^2 (\vel A_1 \ver^2 + \vel C_1 \ver^2)
+ 2 \hat{s} A_1 C_1^\ast \Big] \nnb \\
%2
\ar 6 m_B \hat{m}_\ell \hat{s} \lambda
(B_1-D_1) (B_4^\ast - B_5^\ast) \nnb \\
%3
\ek 16 m_B^5 \hat{m}_\ell \hat{s} \lambda^2
(B_2 + D_2) B_7^\ast C_{TE}^\ast \nnb \\
%4
\ek 6 m_B^2 \hat{m}_\ell^2 \hat{s} \lambda
\Big[ \vel B_4 \ver^2 +  \vel B_5 \ver^2 -
2 (B_1 - D_1) (B_3^\ast - D_3^\ast) \Big] \nnb \\
%5
\ek 6 m_B^3 \hat{m}_\ell \hat{s} \lambda (1-\hat{r}_{\rho})
\Big[ (B_2 - D_2) (B_4^\ast - B_5^\ast) + 2 m_B \hat{m}_\ell
(B_2 - D_2) (B_3^\ast - D_3^\ast) \Big] \nnb \\
%6
\ek 6 m_B^3 \hat{m}_\ell \hat{s}^2 \lambda
(B_3 - D_3) (B_4^\ast - B_5^\ast) \nnb \\
%7
\ek 6 m_B^4 \hat{m}_\ell^2 \hat{s}^2 \lambda
\vel B_3 - D_3 \ver^2 \nnb \\
%8
\ar 16 m_B^3 \hat{m}_\ell \hat{s} \lambda
(1-\hat{r}_{\rho}-\hat{s})
\Big[ (B_1 + D_1) B_7^\ast C_{TE}^\ast +
2 (B_2 + D_2) B_6^\ast C_{TE}^\ast \Big] \nnb \\
%9
\ar 4 m_B^2 \hat{m}_\ell^2 \lambda
(4 - 4 \hat{r}_{\rho} - \hat{s}) (B_1 B_2^\ast + D_1 D_2^\ast) \nnb \\
%10
\ar 2 \hat{s}
[6 \hat{r}_{\rho} \hat{s} (1-v^2) + \lambda (1-3 v^2)]
B_1 D_1^\ast \nnb \\
%11
\ek 2 m_B^4 \hat{m}_\ell^2 \lambda
[\lambda + 3 (1-\hat{r}_{\rho})^2]
(\vel B_2 \ver^2 + \vel D_2 \ver^2) \nnb \\
%12
\ek m_B^2 \hat{s} \lambda
[2 - 2 \hat{r}_{\rho} + \hat{s} - 3 v^2 (2 - 2 \hat{r}_{\rho} - \hat{s})]
(B_1 D_2^\ast + B_2 D_1^\ast) \nnb \\
%13
\ek 8 \hat{m}_\ell^2
(\lambda - 3 \hat{r}_{\rho}\hat{s})
(\vel B_1 \ver^2 + \vel D_1 \ver^2) \nnb \\
%14
\ek 32 m_B \hat{m}_\ell \hat{s}
(\lambda - 12 \hat{r}_{\rho}\hat{s})
(B_1 + D_1) B_6^\ast C_{TE}^\ast \nnb \\
%15
\ek 8 m_B^4 \hat{s}^2 \lambda
\Big[ \lambda m_B^2 \vel B_7 \ver^2 -
4 (1-\hat{r}_{\rho} - \hat{s}) B_6 B_7^\ast \Big]
\Big[ v^2 \vel C_T \ver^2 + 4 (1-2 v^2) \vel C_{TE} \ver^2 \Big] \nnb \\
%16
\ar 3 m_B^2 \hat{s}^2 \lambda (1+v^2) B_4 B_5^\ast \nnb \\
%17
\ek m_B^4 \hat{s} \lambda
\Big[ (1+3 v^2) \lambda - 3 (1-v^2) (1-\hat{r}_{\rho})^2 \Big]
B_2 D_2^\ast \nnb \\
%18
\ek 32 m_B^2 \hat{s}^2
\Big\{ (\lambda + 12 \hat{r}_{\rho} \hat{s}) v^2 \vel C_T \ver^2 +
4 [ \lambda (1-2 v^2) -  12 \hat{r}_{\rho} \hat{s}]
\vel C_{TE} \ver^2 \Big\} \vel B_6 \ver^2 \nnb \\
%19
\ek 128 m_B^2
\Big\{ 4 \lambda [\lambda - (1-\hat{r}_{\rho})^2] +
8 \hat{s} (1-\hat{r}_{\rho})
[\lambda - 6 \hat{r}_{\rho} (1-\hat{r}_{\rho})] +
8 \lambda \hat{s} v^2 (8 \hat{r}_{\rho} - \hat{s}) \Big\}
\vel C_{TE} \ver^2 \vel t_1 \ver^2 \nnb \\
%20
\ar 128 m_B^2
\Big\{ 16 \lambda \hat{r}_{\rho} \hat{s} -
\lambda [\lambda - (1-\hat{r}_{\rho})^2] v^2 -
2 \hat{s} v^2
[\lambda (1+3 \hat{r}_{\rho}) + 6 \hat{r}_{\rho} (1-\hat{r}_{\rho})^2]\Big\}
\vel C_T \ver^2 \vel t_1 \ver^2 \nnb \\
%21
\ar 128 m_B^2 \hat{s}^2
\Big\{ [\lambda + 12 \hat{r}_{\rho} (1-\hat{r}_{\rho})]
v^2 \vel C_T \ver^2 +
4 [\lambda(1-2 v^2) - 12 \hat{r}_{\rho} (1-\hat{r}_{\rho})]\Big\}
B_6 t_1^\ast \nnb \\
%22
\ek 64 m_B^4 \hat{s}^2 \lambda  
(1+ 3 \hat{r}_{\rho} - \hat{s})
\Big[ v^2 \vel C_T \ver^2 + 4 (1-2 v^2) \vel C_{TE} \ver^2 \Big]
B_7 t_1^\ast \nnb \\
%23
\ar 64 m_B \hat{m}_\ell \hat{s}
[\lambda - 12 \hat{r}_{\rho} (1-\hat{r}_{\rho})]
(B_1 + D_1) C_{TE}^\ast t_1^\ast \nnb \\
%24
\ar 512 m_B^3 \hat{m}_\ell \hat{r}_{\rho} \hat{s} \lambda
(A_1 + C_1) C_T^\ast t_1^\ast \nnb \\
%25
\ek 64 m_B^3 \hat{m}_\ell \hat{s} \lambda
(1+ 3 \hat{r}_{\rho} - \hat{s})
(B_2 + D_2) C_{TE}^\ast t_1^\ast
\bigg\}~,
\eea

\section{Numerical analysis}    

In this section we analyze the effects of the Wilson coefficients on
the polarized $FB$ asymmetry. The input parameters we use in our numerical
calculations are: $m_{\rho}=0.77~GeV$, $m_{\tau}=1.77~GeV$,
$m_{\mu}=0.106~GeV$, $m_{b}=4.8~GeV$, $m_{B}=5.26~GeV$ and 
$\Gamma_B = 4.22\times 10^{-13}~GeV$. 
For the values of the Wilson coefficients we use
$C_7^{SM}=-0.313,~C_9^{SM}=4.344$ and $C_{10}^{SM}=-4.669$. It should be
noted that the above--presented value for $C_9^{SM}$ corresponds only to
short distance contributions. In addition to the short distance
contributions, it receives long distance contributions which result from 
the conversion of $\bar{u}u$, $\bar{d}d$ and $\bar{c}c$ to the lepton pair. 
In order to minimize the hadronic uncertainties we will discard the regions
around low lying resonances $\rho$, $w$, $J/\psi$, $\psi^\prime$, 
$\psi^{\prime\prime}$, by dividing the $q^2$
region to low and high dilepton mass intervals:
\bea
\begin{array}{ll}
\mbox{\rm Region I:}& 1~GeV^2 \le q^2 \le 8~GeV^2~,\\
\mbox{\rm Region II:}& 14.5~GeV^2 \le q^2 \le (m_B-m_\rho)^2~GeV^2~,
\end{array} \nnb
\eea
where the contributions of the higher $\psi$ resonances do still exist in
the second region.
The form factors we have used in the present work are more refined ones
predicted by the light cone QCD sum rules \cite{R7024}.
The $q^2$ dependence of the form factors for the $B \rar \rho$ transition 
can be represented in the following form:

\bea
\label{e7028}
F(q^2)\es\frac{r_1}{1-q^2/m_{\rm res}^2}+\frac{r_2}{1-q^2/m_{\rm fit}^2}~, \\
\label{e7029}
F(q^2)\es \frac{r_2}{1-q^2/m_{\rm fit}^2}~, \\
\label{e7030}
F(q^2)\es\frac{r_1}{1-q^2/m_{\rm fit}^2}+\frac{r_2}{(1-q^2/m_{\rm
fit}^2)^2}~,
\eea
with the three independent parameters $r_1$, $r_2$ and $m_{\rm fit}$ 
being listed listed in Table 1. The dominant poles at $q^2=m_{\rm res}^2$ 
correspond to the resonances
\bea
J^P = \left\{ \begin{array}{ll}
1^- & \mbox{\rm for $V$}~,\\
0^- & \mbox{\rm for $A_0$}~,\\
1^+ & \mbox{\rm for $A_1,A_2,A_3$ and $T_2,T_3$}~.\end{array} \right. \nnb
\eea
The values of the parameters $r_1$, $r_2$ and $m_{\rm fit}$ for various form
factors are presented in Table-1.
\begin{table}[tbh]
\renewcommand{\arraystretch}{1}
\addtolength{\arraycolsep}{3pt}
\vskip-10pt
$$
\begin{array}{|l||cccc|c|}
\hline
& r_1 & m_{\rm res}^2~(GeV^2) & r_2 &
m_{\rm fit}^2~(GeV^2) &  \mbox{\rm Fit Eq.} \\
\hline
V^{B_q \to \rho} & \phantom{-}1.045 & 5.32^2 &
-0.721 & 38.34 & (\ref{e7028}) \\ 
A_0^{B_q \to \rho} & \phantom{-}1.527 & 5.28^2 & -1.220  & 33.36 & (\ref{e7028}) \\ 
A_1^{B_q \to \rho} & \phantom{-} -     & -         & \phantom{-}0.240 & 37.51 & (\ref{e7029}) \\ 
A_2^{B_q \to \rho} & \phantom{-}0.009 & -         &
\phantom{-}0.212 & 40.82 & (\ref{e7030}) \\ 
T_1^{B_q \to \rho} & \phantom{-}0.897 & 5.32^2 & -0.629 & 38.04 & (\ref{e7028}) \\ 
T_2^{B_q \to \rho} & \phantom{-} -     & -         & \phantom{-}0.267 & 38.59 & (\ref{e7029}) \\ 
\widetilde{T}_3^{B_q \to \rho} & \phantom{-}0.022 & -         & \phantom{-}0.246 & 40.88 & (\ref{e7030}) \\
\hline
\end{array}
$$
\vskip-1pt
\addtolength{\arraycolsep}{-3pt}
\caption{$B \rar \rho$ decay form factors in a three-parameter 
$r_1$, $r_2$ and $m_{\rm fit}$ fit.}
\renewcommand{\arraystretch}{1}
\addtolength{\arraycolsep}{-3pt}
\end{table}

Note that $T_3$ entering into Eqs. (\ref{e7009}) and (\ref{e7010})
is related to $\widetilde{T}_3$ as follows::
\bea
T_3 = \frac{m_B^2-m_\rho^2}{q^2} (\widetilde{T}_3 - T_2)~. \nnb
\eea 

In the numerical analysis the values of the new Wilson coefficients
which describe the new physics beyond the SM are needed. In our
calculations the new Wilson coefficients are varied in the range $-\ve
C_{10}^{SM}\ve \le \ve C_X\ve \le \ve C_{10}^{SM}\ve$. The experimental
results on the branching ratio of the $B \rar K^\ast(K) \ell^+ \ell^-$ 
decay \cite{R7025,R7026} and the upper limit on the branching ratio of  $B
\rar \mu^+ \mu^-$ \cite{R7027} suggests that that this is the right order of
magnitude for the new Wilson coefficients.

It follows from the expressions of all nine double--lepton polarization
asymmetries that depend both on $q^2$ and the new Wilson coefficients $C_X$. 
Therefore, it may experimentally be difficult
to study these dependencies at the same time. For this reason, we eliminate
$q^2$ dependence by performing integration over $q^2$ in the allowed region,
i.e., we consider the averaged double--lepton polarization asymmetries. The
averaging over $q^2$ is defined as
\bea
\la P_{ij} \ra = \frac{\ds \int_{R_i}
P_{ij} \frac{d{\cal B}}{d \hat{s}} d \hat{s}}
{\ds \int_{R_i}
\frac{d{\cal B}}{d \hat{s}} d \hat{s}}~,\nnb
\eea
where $R_i=$ Regions I or II, over which the integrations are calculated.
We present our analysis in a series of figures.

In Figs. (1) and (2) we present the dependence of $\lla P_{LL} \rra$ on
$C_X$ for the $B \rar \rho \mu^+ \mu^-$ decay in the regions I and II,
respectively. The intersection of all curves corresponds to the SM case.
From these figures we see that $\lla P_{LL} \rra$ exhibits strong dependence
only on the tensor interactions $C_T$ and $C_{TE}$, and has practically
symmetric behavior in regard to its dependence on $C_T$ and $C_{TE}$ with
respect to zero position. Furthermore, $\lla P_{LL} \rra$ seems to be
independent of all remaining new Wilson coefficients.

We depict from Figs. (3) and (4) the dependence of $\lla P_{LT} \rra$ on
$C_X$ for the $B \rar \rho \mu^+ \mu^-$ decay in the regions I and II,
respectively.We observe from these figures that $\lla P_{LT} \rra$ is
sensitive to to the existence of scalar  $C_{LRLR},C_{RLLR}$ and tensor
interactions $C_T,C_{TE}$ and it shows weak dependence on all remaining
coefficients. A striking feature of its behavior is that $\lla P_{LT} \rra$
changes its sign in the above--mentioned region of the new Wilson
coefficients, while in the SM case its sign never changes. For this reason
study of the magnitude and sign of $\lla P_{LT} \rra$ can serve as a good
test for looking new physics beyond the SM.  

The dependence of $\lla P_{TL} \rra$ on $C_X$ for the $B \rar \rho \mu^+
\mu^-$ decay is presented in Fig. (5) in
Region I and Fig. (6) in Region II, respectively. In both regions $\lla
P_{TL} \rra$ exhibits strong dependence on scalar $C_{RLRL}$ and $C_{LRRL}$
and tensor interaction coefficients. Moreover, when $C_{RLRL}(C_{LRRL})$ is
negative (positive), $\lla P_{TL} \rra$ is positive (negative). When
$C_{TE}<-0.8 (>0)$ and $C_T<0 (>2)$, $\lla P_{TL} \rra$ is negative and
positive otherwise. Hence determination of the magnitude and sign of $\lla
P_{TL} \rra$ gives unambiguous confirmation of the existence of new physics
due to scalar and tensor interactions.

In Figs. (7) and (8) we present the dependence of $\lla P_{TT} \rra$  
on $C_X$ for the $B \rar \rho \mu^+ \mu^-$ decay. In region I (see Fig. (7))
$\lla P_{TT} \rra$ is strongly dependent on vector type interactions
$C_{LR},C_{RR}$ and for the negative values of $C_{LL}$ and $C_T$. On the
other hand, in Region II, $\lla P_{TT} \rra$ is strongly dependent only on
tensor interaction. In Region I $\lla P_{TT} \rra$ is positive (negative)
for negative values of $C_{LR}(C_{RR})$ and it attains at negative
(positive) values for positive values of $C_{LR}(C_{RR})$. In the second
region the sign of $\lla P_{TT} \rra$ changes only for the vector
interaction $C_{RR}$.

Depicted in Figs. (9) and (10) are the dependence of $\lla P_{NN} \rra$ on 
the new Wilson coefficients. The situation is quite similar to the previous
case for the $\lla P_{TT} \rra$. The only difference being, $\lla P_{NN}
\rra$ in Region I depends strongly on $C_{TE}$ rather than $C_T$, for their
negative values, compared to that for the $\lla P_{TT} \rra$ case.

All remaining double--lepton polarization asymmetries for the $B \rar \rho
\mu^+ \mu^-$ decay are very small numerically and therefore we do not
present them.

Through Figs. (11)--(14) we study the dependence of $P_{ij}$ on the new Wilson
coefficients for the $B \rar \rho \tau^+ \tau^-$ decay, which provides
richer information about the new physics effects.

In Fig. (11) the dependence of $\lla P_{LL} \rra$ on $C_X$ is given. We
observe from this figure that $\lla P_{LL} \rra$ is very sensitive to all
new Wilson coefficients except $C_{RL}$. It changes its sign
only for the variations in $C_T$ and for all rest of the new Wilson coefficients
$\lla P_{LL} \rra$ does not seem to change its sign. Therefore 
investigation of the sign of $\lla P_{LL} \rra$ can give important clue
about the existence of the tensor interaction.

In Fig. (12) we present the dependence of of $\lla P_{LT} \rra$ on the new
Wilson coefficients. Noting that $\lla P_{TL} \rra$ exhibits similar
behavior, except several scalar coefficients, $\lla P_{LT} \rra$
is sensitive to all remaining Wilson coefficients. Similar to the $\lla
P_{LL} \rra$ case, $\lla P_{LT} \rra$ changes its sign in the presence of
the tensor interaction and therefore this circumstance can be quite useful
in looking for new physics beyond the SM. 

The dependence of $\lla P_{LN} \rra \approx - \lla P_{NL} \rra$ on $C_X$ is
presented in Fig. (13). We see from this figure that $\lla P_{LN} \rra$ is
very sensitive to all new Wilson coefficients, especially to the vector
interaction coefficients $C_{LL}$ and $C_{LR}$.

In Fig. (14) we present the dependence of $\lla P_{NN} \rra \approx - \lla
P_{TT} \rra$ on the new Wilson coefficients. We observe from this figure
that when $C_X$ is negative $\lla P_{NN} \rra > \lla P_{NN}^{SM} \rra$ for
the coefficients $C_{LR}$, $C_{LL}$, $C_{LRRL}$ and $C_T$, and $\lla P_{NN}
\rra > \lla P_{NN}^{SM} \rra$ for  the coefficients $C_{RL}$, $C_{RR}$, 
$C_{RLLR}$ and $C_{TE}$. On the other hand, when $C_X$ is positive the
situation changes to the contrary, except for the tensor interaction
(neglecting the narrow region for the coefficient $C_{TE}$). The numerical
analysis for the rest of the remaining double--lepton polarization 
asymmetries for the $B \rar \rho \tau^+ \tau^-$ decay are not presented in
this work due to their negligible smallness.     

It follows from the present analysis that few of the double--lepton 
polarization asymmetries show considerable departure from the SM 
predictions and these ones are strongly dependent on different types of 
interactions. Hence, the study of these quantities can play crucial role 
in establishing new physics beyond the SM.       
 
At the end of this section, we would like to discuss the following problem.
Could there be a case in which the branching ratio coincides
with that of the SM result, while double--lepton polarization asymmetry does
not? In order to answer this question we study the correlation between the
$\lla P_{ij} \rra$ and the branching ratio ${\cal B}$. We can
briefly summarize the results of our numerical analysis as follows:
For the $B\rar \rho \mu^+ \mu^-$ decay, except for a very 
narrow region of $C_{RR}$, such a region is absent
for all new Wilson coefficients for all of the asymmetries $\lla P_{ij}
\rra$.

The $B\rar \rho \tau^+ \tau^-$ decay is more informative for this aim, which
are measurable in the experiments. In
Figs. (15) and (16) we present the dependence of $\lla P_{LL}
\rra$ and $\lla P_{LT} \rra$ on the branching ratio. It follows
from these figures that, there indeed exists certain regions of $C_X$ for
which the double--lepton polarization asymmetry differs from the SM
prediction, while the branching ratio coincides with that of the SM result.
We also note that, such a region exists for the remaining double--lepton
polarization asymmetries for the tensor interaction as well. 
 
In conclusion, in the present work we investigate the double--lepton
polarization asymmetries when both leptons are polarized, using a general, model
independent form of the effective Hamiltonian. We obtain that various
double--lepton polarization asymmetries can serve as a good test in looking
for new physics beyond the SM. We also study the correlation between $\lla
P_{ij}\rra$ and the branching ratio for the $B\rar \rho \tau^+ \tau^-$ decay
and find out that there exist regions of the new Wilson coefficients for
which the double--lepton polarization asymmetry differs considerably from
the SM prediction, while the branching ratio coincides with the SM
prediction. Therefore in these regions the new physics effects can be
established just by measuring the double--lepton polarizations.

\newpage

\section*{Figure captions}
{\bf Fig. (1)} The dependence of the averaged
double--lepton polarization asymmetry $\lla P_{LL} \rra$
on the new Wilson coefficients $C_X$, for the $B \rar \rho \mu^+ \mu^-$
decay, in Region I.\\ \\
{\bf Fig. (2)} The same as in Fig. (1), but in Region II.\\ \\
{\bf Fig. (3)} The same as in Fig. (1), but for the averaged
double--lepton polarization asymmetry $\lla P_{LT} \rra$.\\ \\
{\bf Fig. (4)} The same as in Fig. (3), but in Region II.\\ \\
{\bf Fig. (5)} The same as in Fig. (1), but for the averaged 
double--lepton polarization asymmetry $\lla P_{TL} \rra$.\\ \\
{\bf Fig. (6)} The same as in Fig. (5), but in Region II.\\ \\
{\bf Fig. (7)} The same as in Fig. (1), but for the averaged
double--lepton polarization asymmetry $\lla P_{TT} \rra$.\\ \\
{\bf Fig. (8)} The same as in Fig. (7), but in Region II.\\ \\
{\bf Fig. (9)} The same as in Fig. (1), but for the averaged
double--lepton polarization asymmetry $\lla P_{NN} \rra$.\\ \\
{\bf Fig. (10)} The same as in Fig. (9), but in Region II.\\ \\
{\bf Fig. (11)} The dependence of the averaged
double--lepton polarization asymmetry $\lla P_{LL} \rra$
on the new Wilson coefficients $C_X$, for the $B \rar \rho \tau^+
\tau^-$ decay, in Region II.\\ \\
{\bf Fig. (12)} The same as in Fig. (11), but for the $\lla P_{LT}\rra$.\\ \\
{\bf Fig. (13)} The same as in Fig. (11), but for the $\lla P_{LN}\rra$.\\ \\
{\bf Fig. (14)} The same as in Fig. (11), but for the $\lla P_{NN}\rra$.\\ \\
{\bf Fig. (15)} Parametric plot of the correlation between the averaged 
double--lepton polarization asymmetry $\lla P_{LL} \rra$ and the 
branching ratio for the $B \rar \rho \tau^+ \tau^-$ decay, 
in Region II.\\ \\
{\bf Fig. (16)} Parametric plot of the correlation between the averaged
double--lepton polarization asymmetry $\lla P_{LT} \rra$ and the 
branching ratio for the $B \rar \rho \tau^+ \tau^-$ decay, 
in Region II.

\newpage

\begin{figure}
\vskip 1.5 cm
    \includegraphics{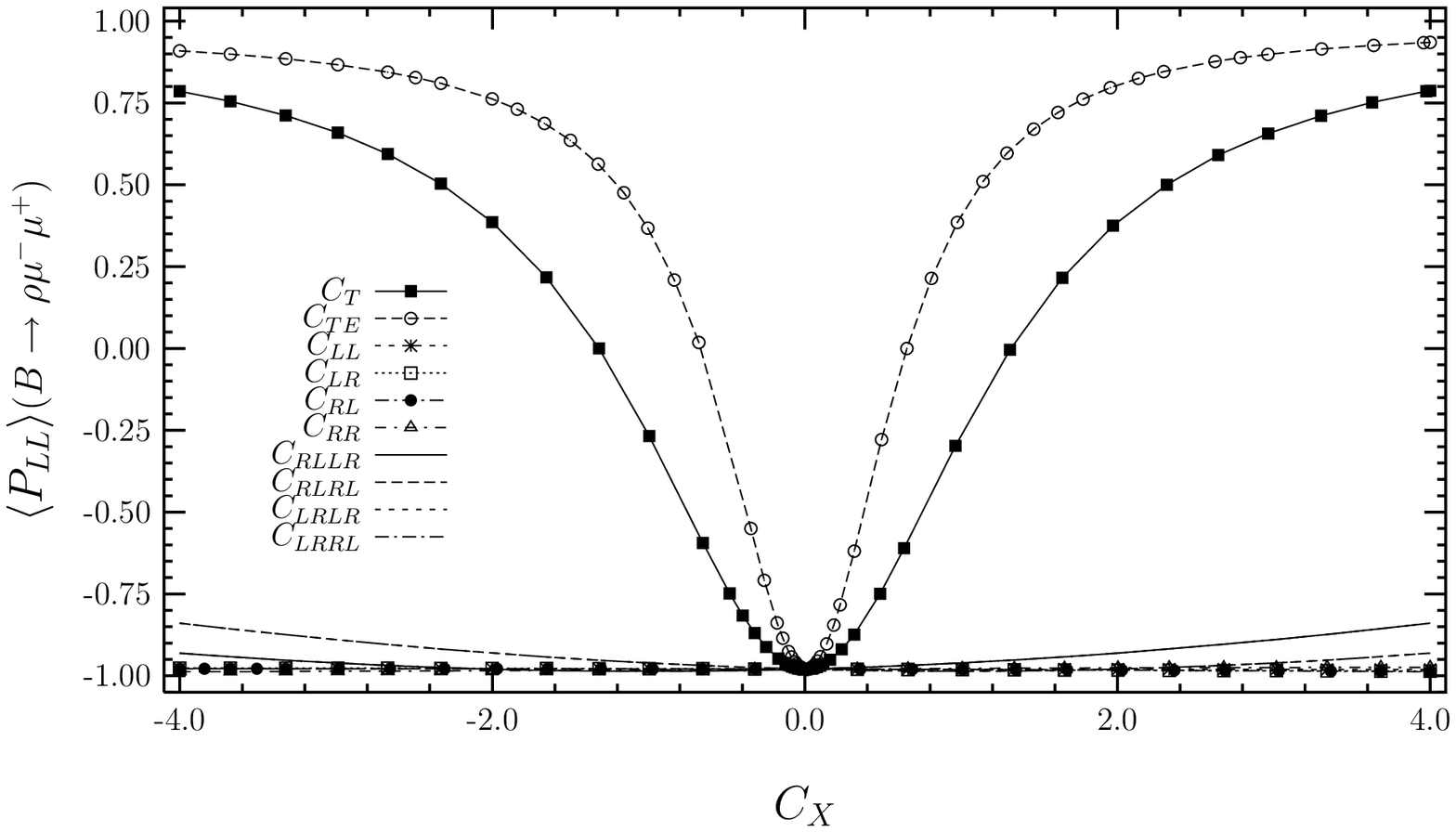}
\vskip 7.8cm
\caption{}
%\begin{center}
%{\bf Fig. 1--a}
%\end{center}
\end{figure}

\begin{figure}
\vskip 2.5 cm
    \includegraphics{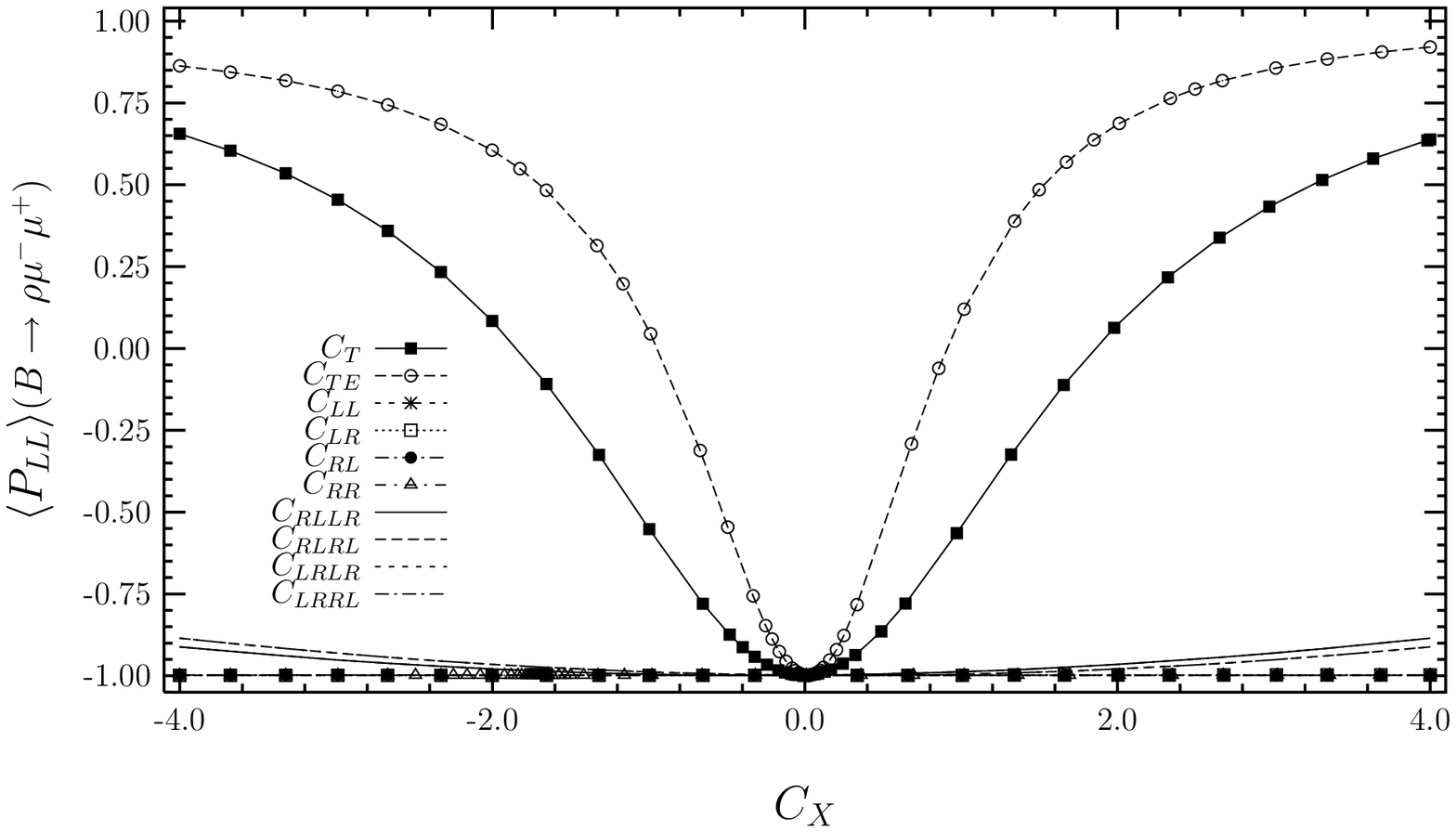}
\vskip 7.8 cm
\caption{}
%\begin{center}
%{\bf Fig. 1--b}
%\end{center}
\end{figure}

\begin{figure}
\vskip 1.5 cm
    \includegraphics{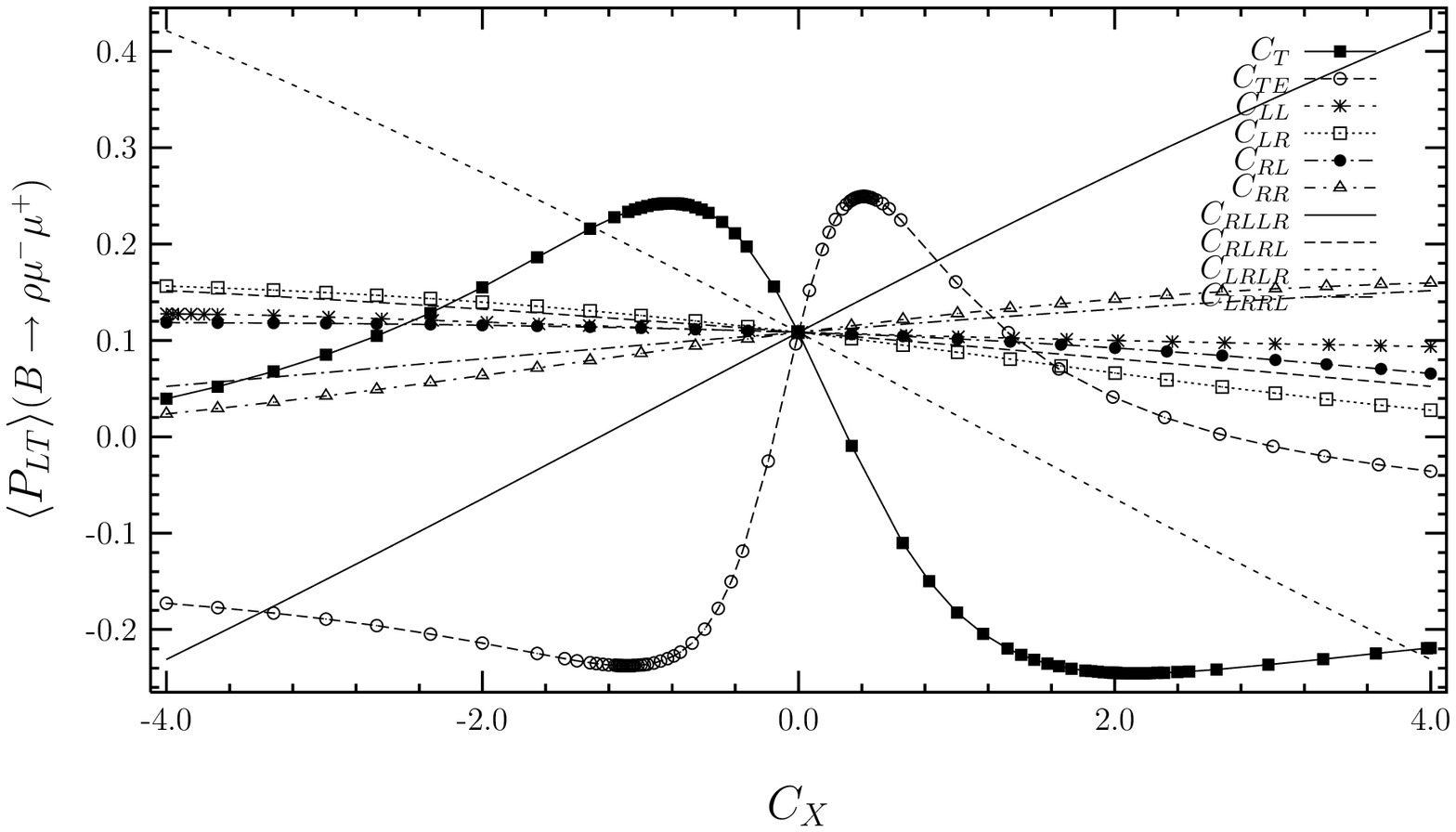}
\vskip 7.8cm
\caption{}
%\begin{center}
%{\bf Fig. 1--a}
%\end{center}
\end{figure}

\begin{figure}
\vskip 2.5 cm
    \includegraphics{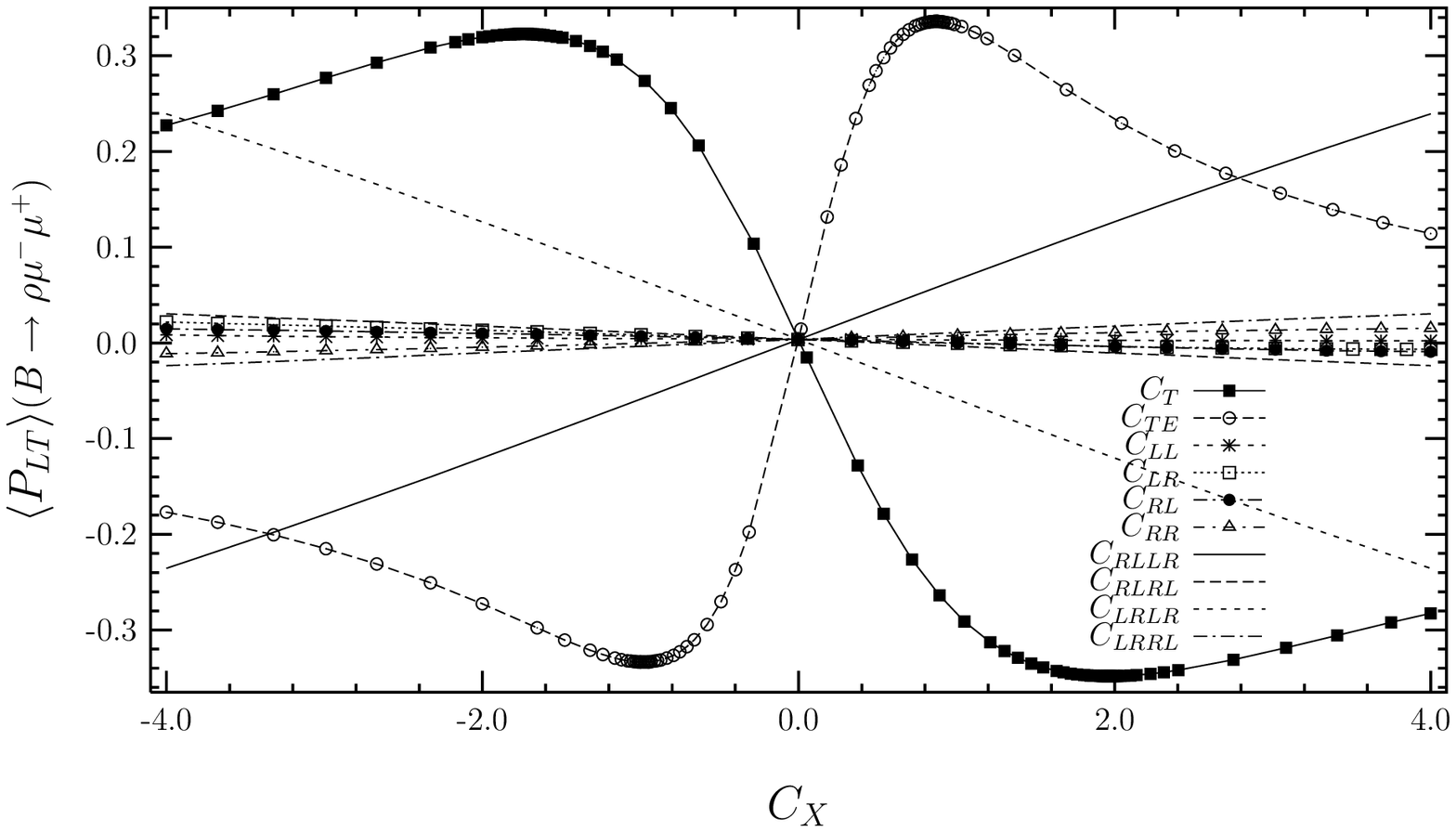}
\vskip 7.8 cm
\caption{}
%\begin{center}
%{\bf Fig. 1--b}
%\end{center}
\end{figure}

\begin{figure}
\vskip 2.5 cm
    \includegraphics{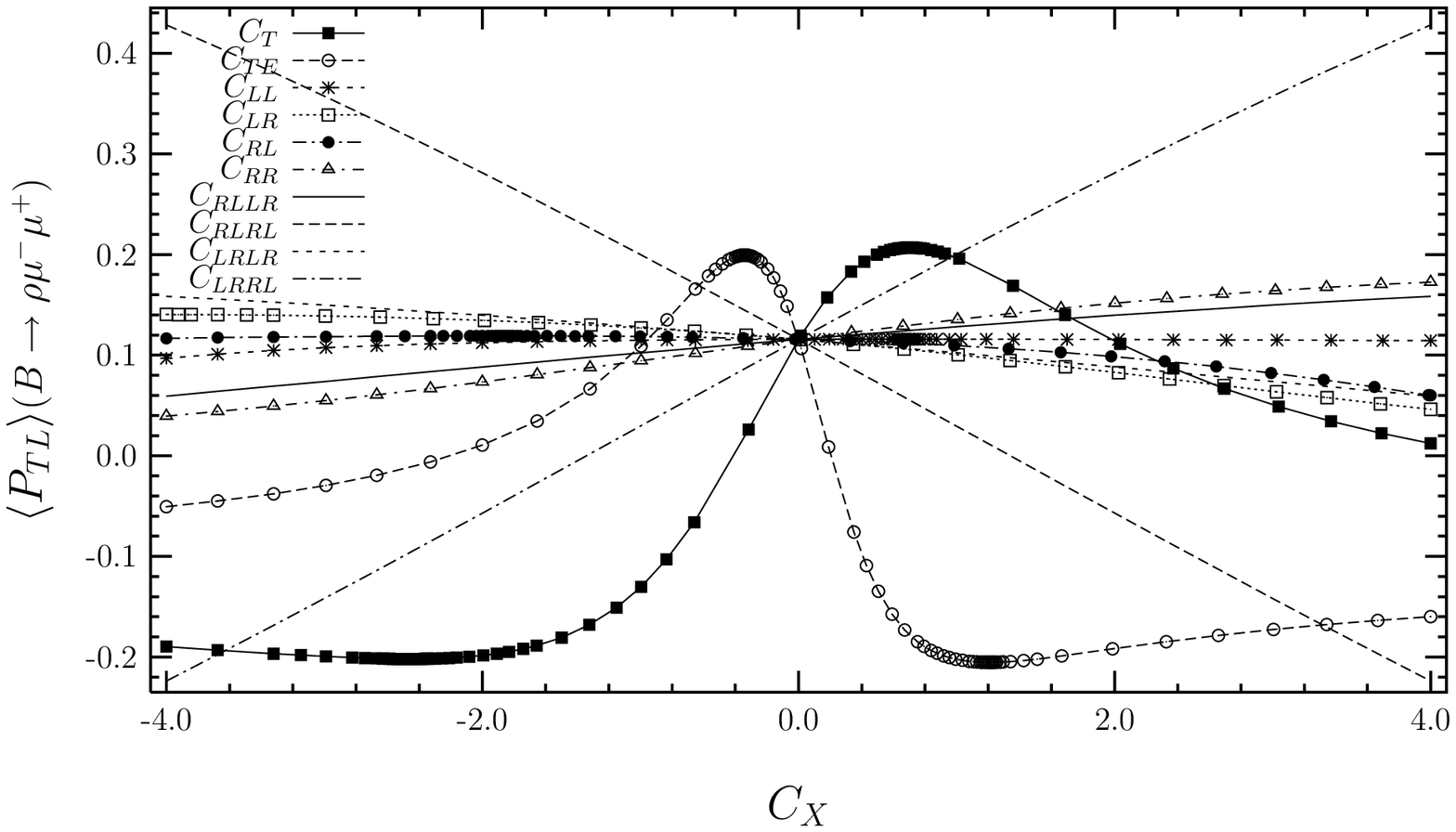}
\vskip 7.8 cm
\caption{}
%\begin{center}
%{\bf Fig. 1--b}
%\end{center}
\end{figure}

\begin{figure}
\vskip 1.5 cm
    \includegraphics{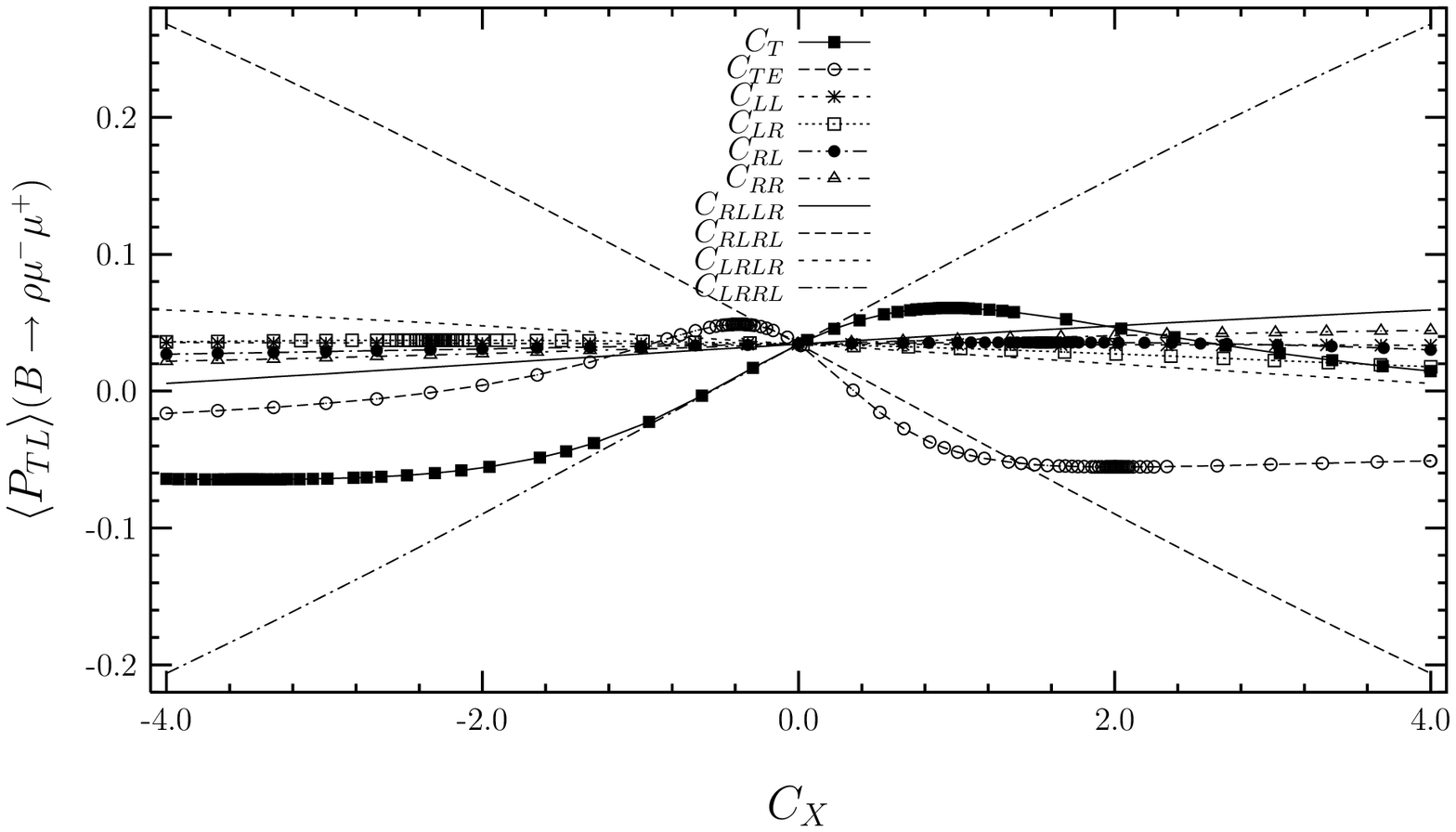}
\vskip 7.8cm
\caption{}
%\begin{center}
%{\bf Fig. 1--a}
%\end{center}
\end{figure}

\begin{figure}
\vskip 2.5 cm
    \includegraphics{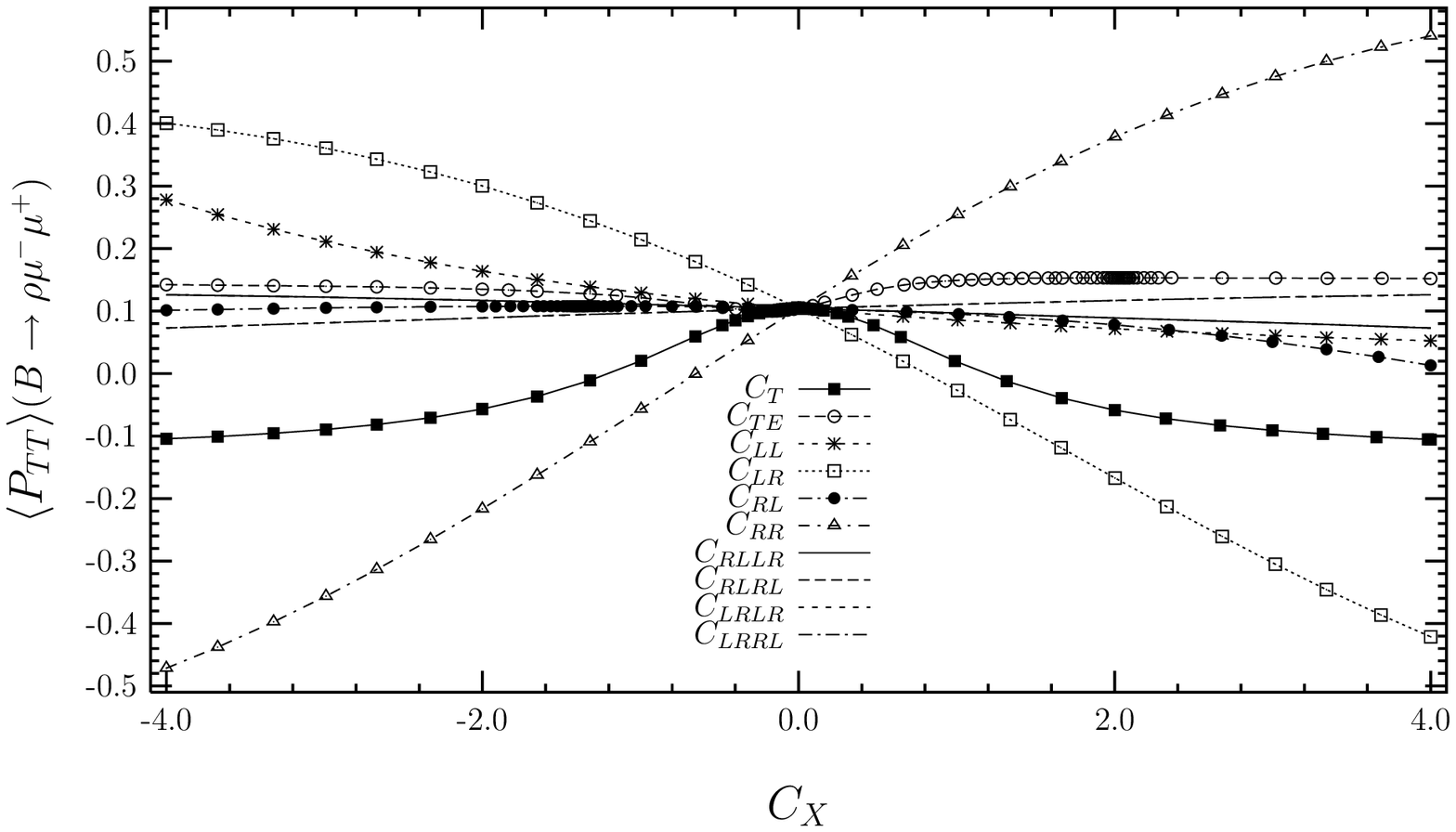}
\vskip 7.8 cm
\caption{}
%\begin{center}
%{\bf Fig. 1--b}
%\end{center}
\end{figure}

\begin{figure}
\vskip 1.5 cm
    \includegraphics{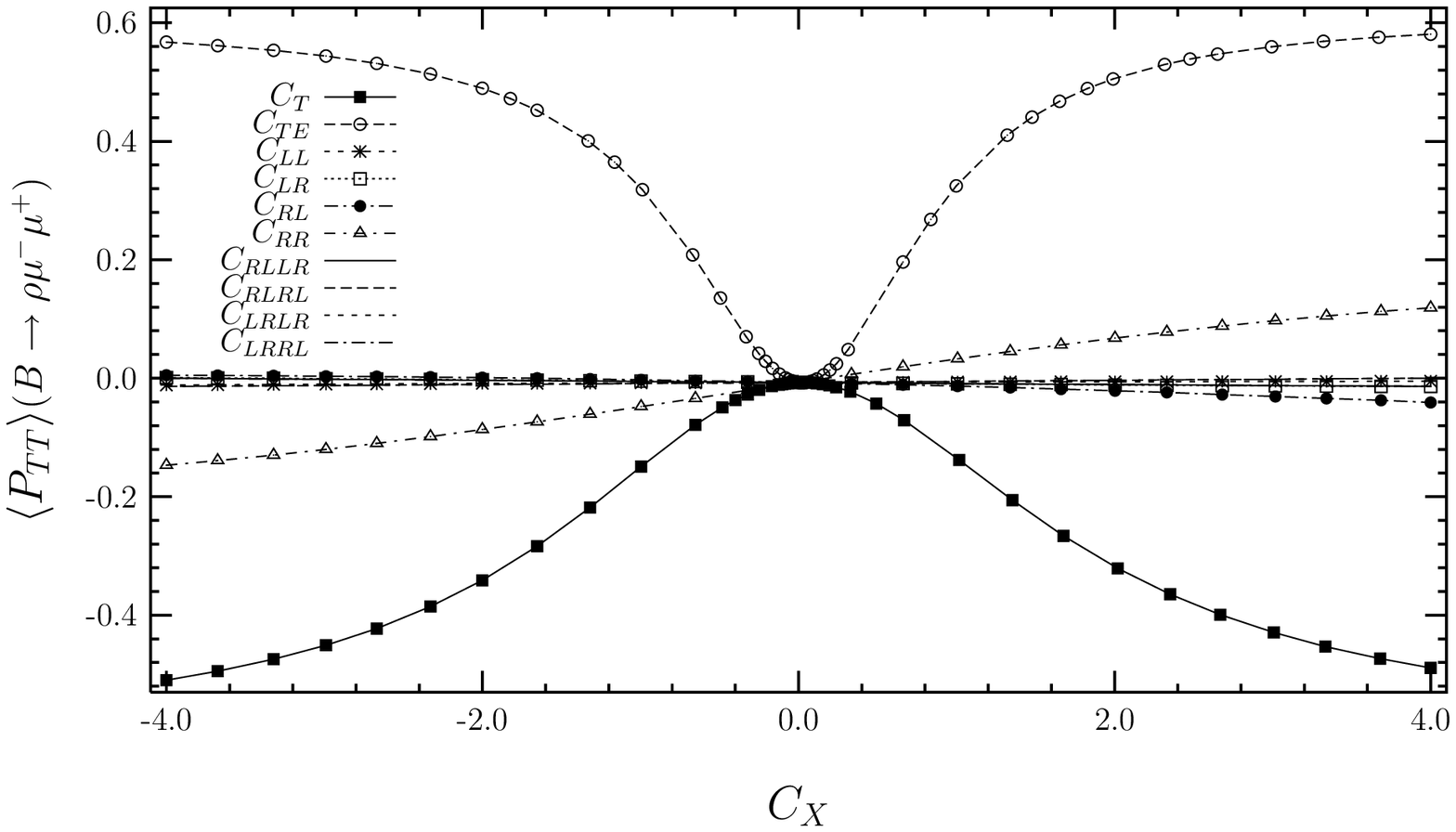}
\vskip 7.8cm
\caption{}
%\begin{center}
%{\bf Fig. 1--a}
%\end{center}
\end{figure}

\begin{figure}
\vskip 2.5 cm
    \includegraphics{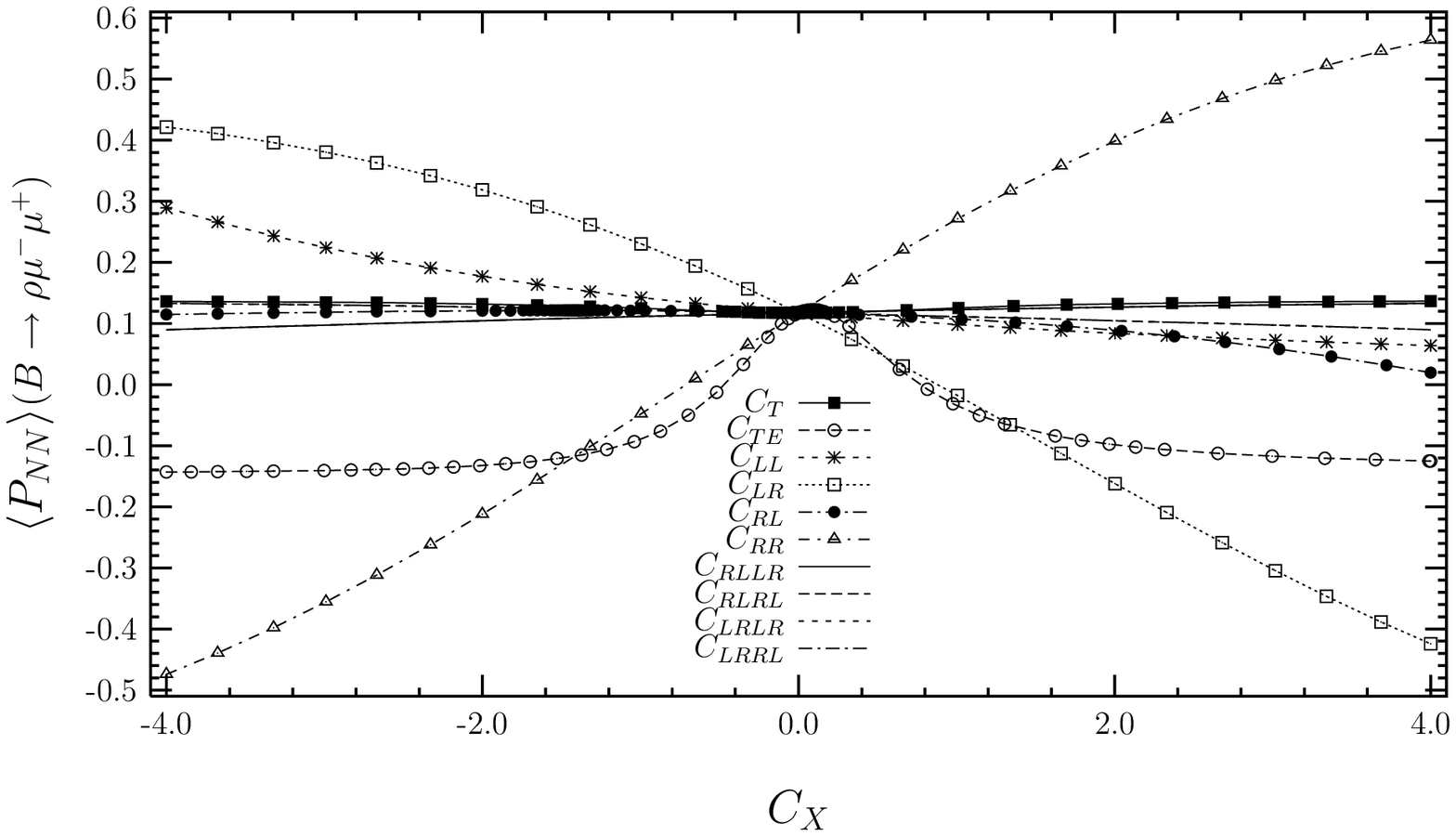}
\vskip 7.8 cm
\caption{}
%\begin{center}
%{\bf Fig. 1--b}
%\end{center}
\end{figure}

\begin{figure}
\vskip 1.5 cm
    \includegraphics{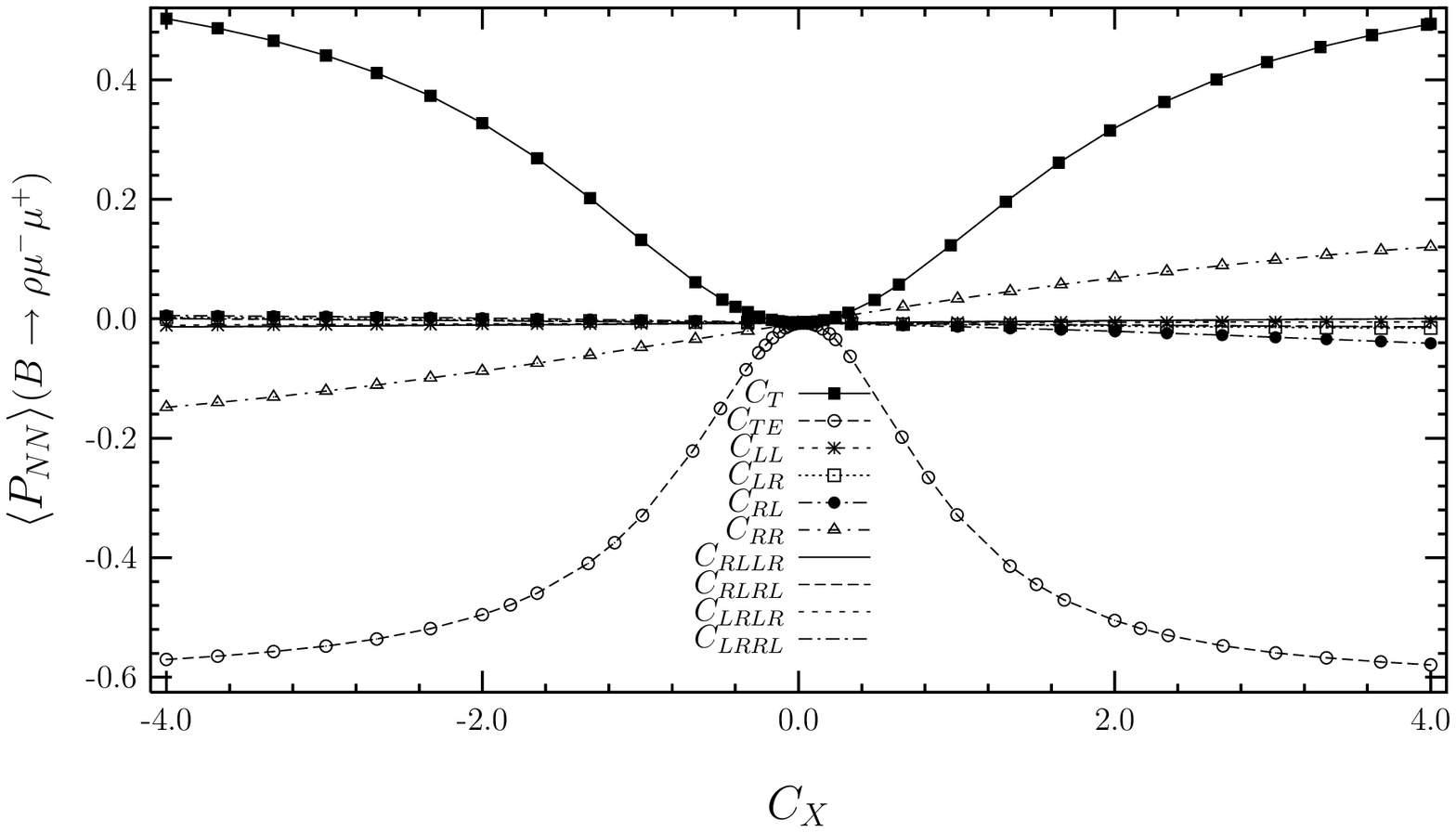}
\vskip 7.8cm
\caption{}
%\begin{center}
%{\bf Fig. 1--a}
%\end{center}
\end{figure}

\begin{figure}
\vskip 2.5 cm
    \includegraphics{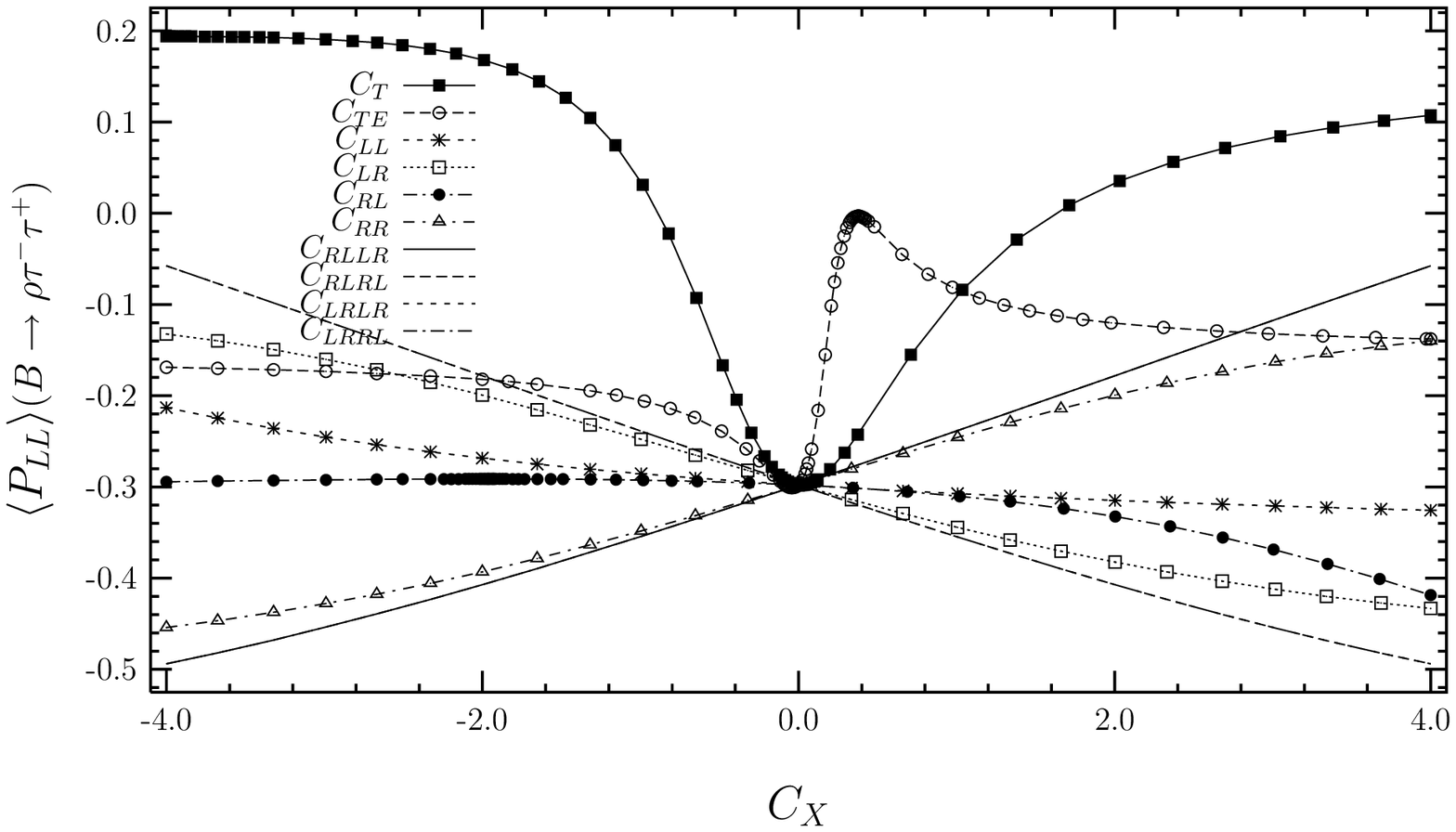}
\vskip 7.8 cm
\caption{}
%\begin{center}
%{\bf Fig. 1--b}
%\end{center}
\end{figure}

\begin{figure}
\vskip 1.5 cm
    \includegraphics{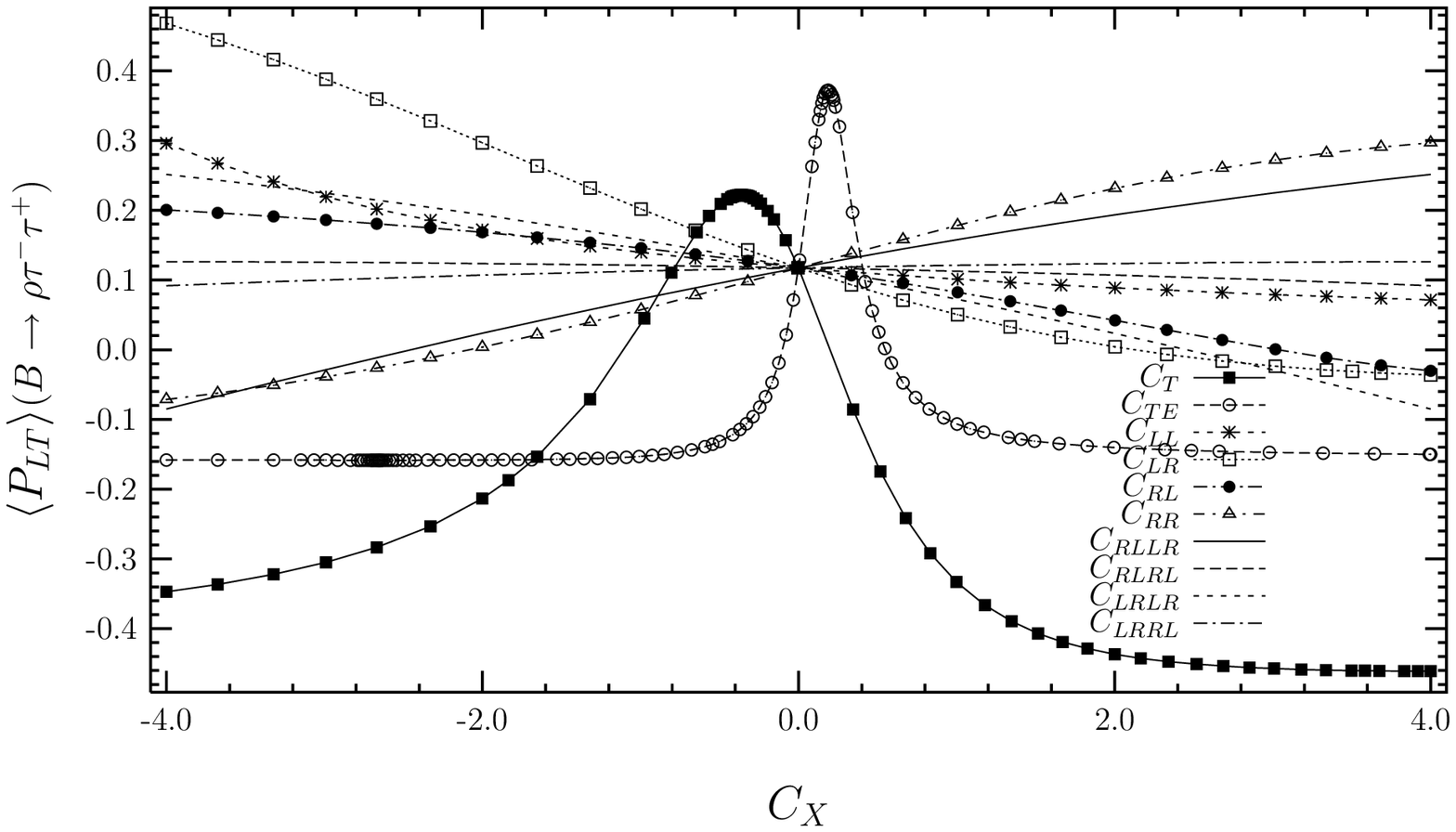}
\vskip 7.8cm
\caption{}
%\begin{center}
%{\bf Fig. 1--a}
%\end{center}
\end{figure}

\begin{figure}
\vskip 2.5 cm
    \includegraphics{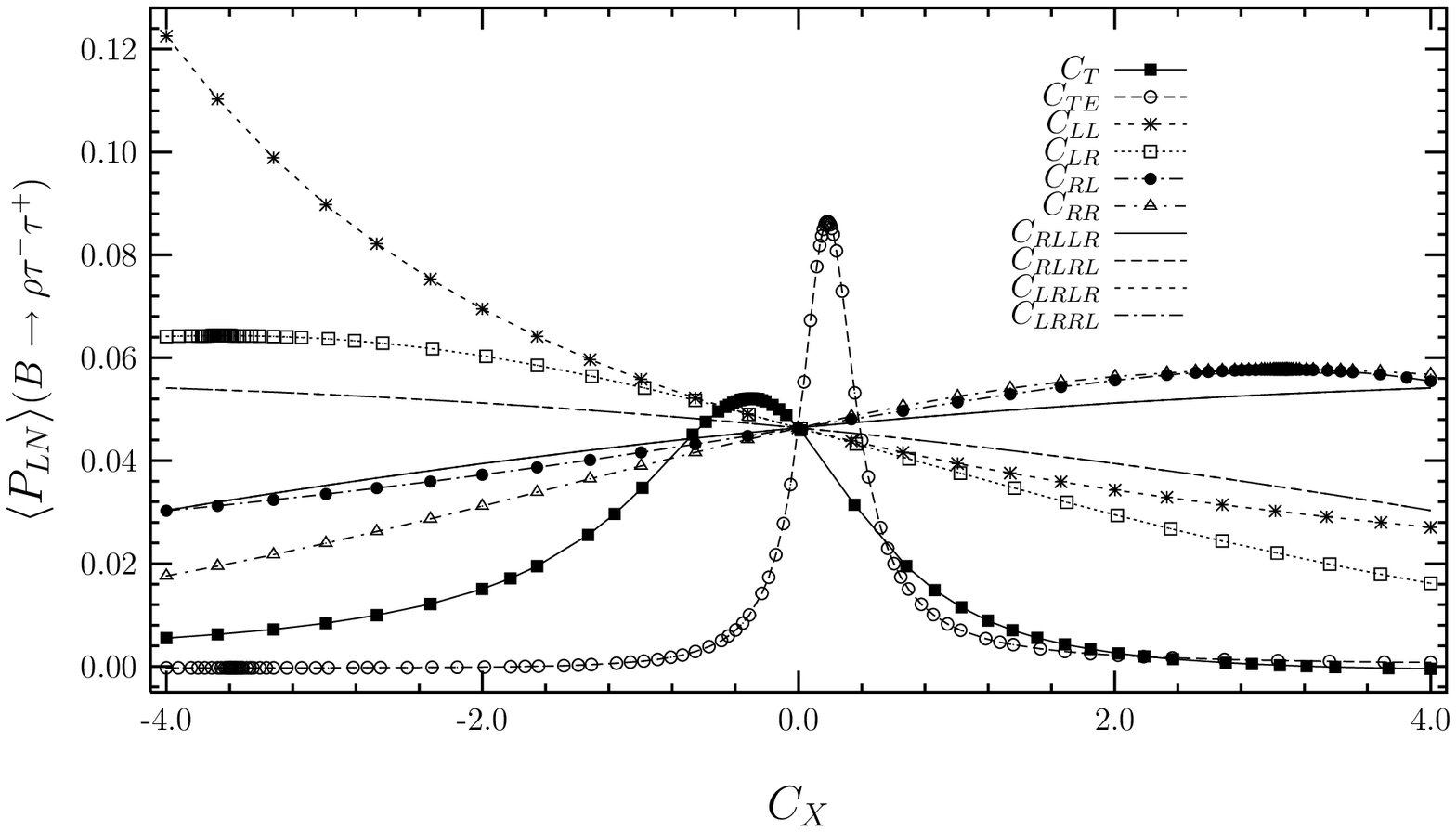}
\vskip 7.8 cm
\caption{}
%\begin{center}
%{\bf Fig. 1--b}
%\end{center}
\end{figure}

\begin{figure}
\vskip 1.5 cm
    \includegraphics{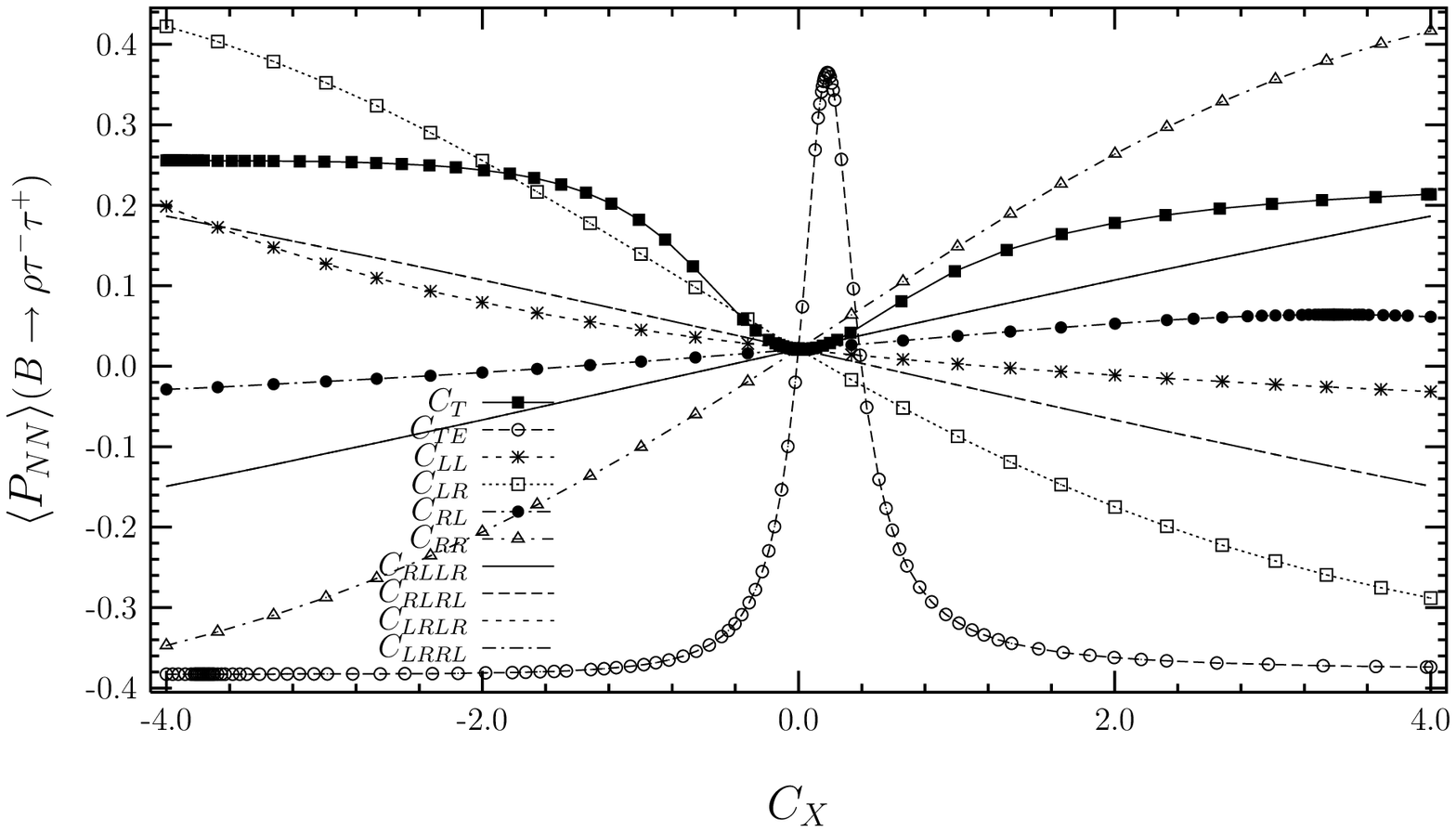}
\vskip 7.8cm
\caption{}
%\begin{center}
%{\bf Fig. 1--a}
%\end{center}
\end{figure}

\begin{figure}
\vskip 2.5 cm
    \includegraphics{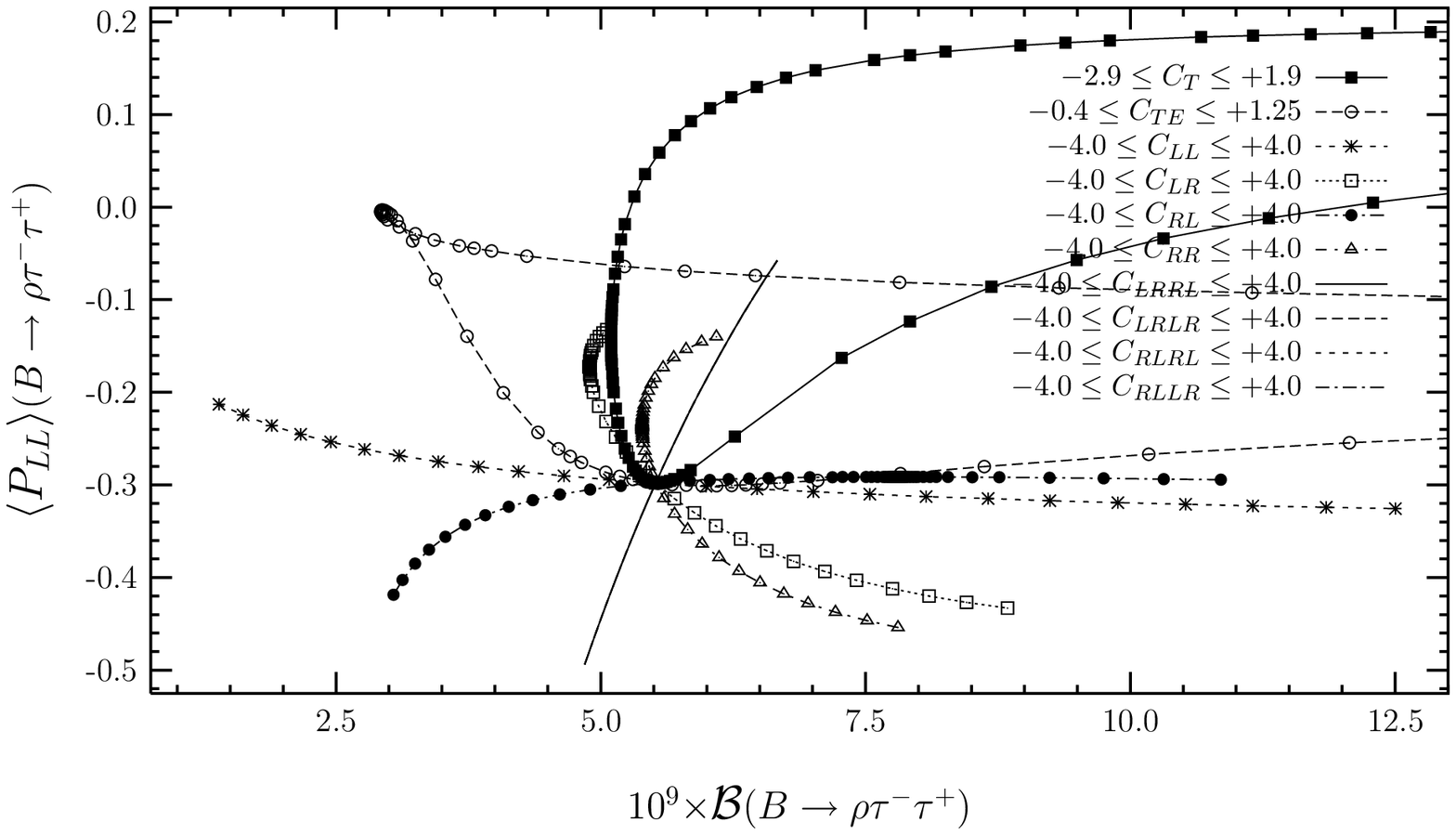}
\vskip 7.8 cm
\caption{}
%\begin{center}
%{\bf Fig. 1--b}
%\end{center}
\end{figure}

\begin{figure}
\vskip 1.5 cm
    \includegraphics{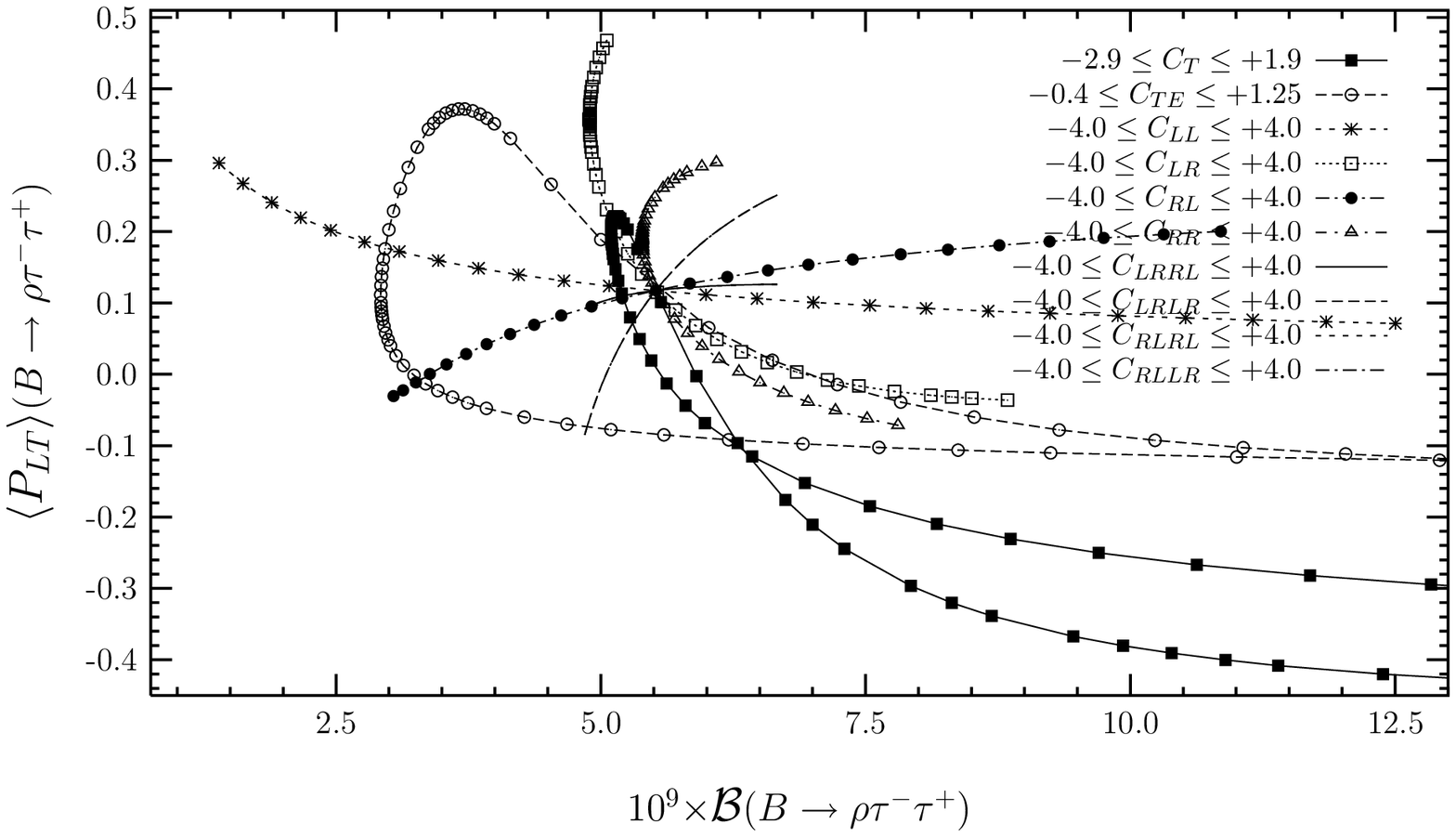}
\vskip 7.8cm
\caption{}
%\begin{center}
%{\bf Fig. 1--a}
%\end{center}
\end{figure}

\end{document}